\documentclass[lettersize,journal]{IEEEtran}
\usepackage{amsmath,amsfonts}
\usepackage{algorithmic}
\usepackage{algorithm}
\usepackage{array}
\usepackage[caption=false,font=normalsize,labelfont=sf,textfont=sf]{subfig}
\usepackage{textcomp}
\usepackage{stfloats}
\usepackage{url}
\usepackage{verbatim}
\usepackage{graphicx}
\usepackage{cite}
\usepackage{todonotes}
\usepackage{hyperref}
\usepackage{cleveref}
\usepackage{siunitx}
\usepackage{xspace}

\newcommand{\eg}{\textit{e.g.,}\xspace}
\newcommand{\ie}{\textit{i.e.,}\xspace}

\newcommand{\R}{\ensuremath{\mathbf{R}}}

\newcommand{\normFro}[1]{\ensuremath{\lVert #1 \rVert_\mathcal{F}}}

\newcommand{\normOmega}[1]{\ensuremath{\lVert #1 \rVert_\Omega}}

\newcommand{\BBP}{\text{BBP443}\xspace}
\newcommand{\KD}{\text{KD490}\xspace}
\newcommand{\CHL}{\text{Chl-a}\xspace}

\definecolor{darkgreen}{rgb}{0.0, 0.45, 0.0}
\definecolor{darkpurple}{rgb}{0.5, 0.0, 0.5}
\definecolor{darkbrown}{rgb}{0.4, 0.26, 0.13}

\begin{document}

\title{Observation-only learning of neural mapping schemes for gappy satellite-derived ocean colour parameters} 



\author{\IEEEauthorblockN{Clément Dorffer, Frédéric Jourdin, Thi Thuy Nga Nguyen, Rodolphe Devillers, David Mouillot and Ronan Fablet}

\thanks{Clément Dorffer, Thi Thuy Nga Nguyen, and Ronan Fablet are with the IMT Atlantique, UMR CNRS Lab-STICC, INRIA team Odyssey, Brest, France}
\thanks{Frédéric Jourdin is with the Service Hydrographique et Océanographique de la Marine (Shom), Brest, France}
\thanks{Rodolphe Devillers is with Espace-Dev (IRD, Univ Montpellier, Univ Guyane, Univ La Réunion, Univ Nouvelle Calédonie, Univ Perpignan Via Domitia), Station SEAS-OI, Saint-Pierre, La Réunion, France}
\thanks{David Mouillot is with MARBEC, University of Montpellier, CNRS, IFREMER, IRD, Montpellier, France}

\thanks{Clément Dorffer completed this work while he was with the IMT-Atlantique, UMR CNRS Lab-STICC, INRIA team Odyssey, Brest, France.}
\thanks{This work was funded by the French ANR, as part of the IA-Biodiv Challenge: Fish-Predict and the Oceanix project.}
\thanks{It was benefited from HPC and GPU resources from GENCI-IDRIS (Grant
2021-101030). And it was received the support from The Regional Council of
Brittany and FEDER through the access of CPER AIDA GPU cluster.}
\thanks{It was also funded by EU Horizon Europe project EDITO Model Lab (Grant
101093293).}
\thanks{This study has been conducted using E.U. Copernicus Marine Service Information; https://doi.org/10.48670/moi-00299.}
\thanks{Corresponding author: Thi Thuy Nga Nguyen (email: nga.nguyen@imt-atlantique.fr).}
}

\markboth{Journal of \LaTeX\ Class Files,~Vol.~14, No.~8, August~2021}%
{Shell \MakeLowercase{\textit{et al.}}: A Sample Article Using IEEEtran.cls for IEEE Journals}


\maketitle

\begin{abstract}

Monitoring optical properties of coastal and open ocean waters is crucial to assessing the health of marine ecosystems. Deep learning offers a promising approach to address these ecosystem dynamics, especially in scenarios where gap-free ground-truth data is lacking, which poses a challenge for designing effective training frameworks. 
Using an advanced neural variational data assimilation scheme (called 4DVarNet), we introduce a comprehensive training framework designed to effectively train directly on gappy data sets. Using the Mediterranean Sea as a case study, our experiments not only highlight the high performance of the chosen neural network in reconstructing gap-free images from gappy datasets but also demonstrate its superior performance over state-of-the-art algorithms such as DInEOF and Direct Inversion, whether using CNN or UNet architectures. 


\end{abstract}

\begin{IEEEkeywords}
space-time interpolation;  data-driven model; data assimilation; image gap filling; observing system experiment (OSE); ocean colour remote sensing; end-to-end deep learning; bio-optical parameter estimation; deep learning in satellite imagery.

\end{IEEEkeywords}


\section{Introduction}
\IEEEPARstart{W}{ater} optical properties are key parameters in understanding and monitoring ocean biogeochemistry and its dynamics \cite{mcclain2009decade}, in particular in the long term \cite{dutkiewicz2019ocean}, including the fate of particles and dissolved material from the continents \cite{gohin2020satellite}. These optical properties allow measurement of phytoplankton biomass and bulk concentration of suspended matter in the water \cite{Bisson2020bbp}. The water transparency, which can be affected by human activities and reveals the amount of light accessible at a given depth available for ocean primary production \cite{mohseni2022ocean}, can also be monitored.

Measuring water optical properties is challenging and requires multiple and repeated observations to get a good representation of the signal, especially within coastal areas where turbidity is highly fluctuating.
While some optical parameters are traditionally measured \textit{in-situ} \cite{bricaud1995situ}, satellite multispectral imaging can provide high-resolution optical observations that can be used over regions of various sizes. For example, the Moderate Resolution Imaging Spectroradiometer (MODIS) sensors \cite{barnes2003status} provide images every 16 days at 300~m to 1~km spatial resolutions for 36 spectral bands (wavelengths: 405 to 14385~nm), while the Ocean and Land Color Instrument (OLCI) sensors \cite{donlon2012global} provide images every 27 days at 300m spatial resolution, for 21 spectral bands (400 to 1020~nm). 
Once pre-processed (e.g., atmospheric corrections), different seawater parameters can be derived from these multispectral data\cite{Hieronymi_2023}. For instance, standard routines\cite{lee_2002,berthon_2004,Dicicco_2017,Volpe_2019}  can be used to estimate Chlorophyll-a (\CHL) concentrations, diffuse attenuation coefficient of light at wavelength 490~nm (\KD), the particulate backscattering coefficient at wavelength 443~nm (\BBP). 
In this article, we selected the \BBP as our variable of interest for its importance in marine ecology and the carbon cycle \cite{loisel2001seasonal}: \BBP has notably been used to assess particulate organic carbon or phytoplankton carbon biomass \cite{Bisson2020bbp, qiu2021relationships}, in Case 1 waters\footnote{In Case-1 waters \cite{morel1977analysis} all bio-optical parameter values recorded at the same time and at the same location depend mainly on a unique variable: Chla. These waters are controlled by biology, which covers large parts of the ocean, mainly located in the open sea.}. In Case 2 waters\footnote{By contrast, Case-2 waters \cite{morel1977analysis} of typically coastal seas are significantly influenced by other constituents such as lithogenic particles and dissolved material typically brought in by river discharges or resuspended from the floor in some coastal areas. Case-2 waters are optically more complex and do not display the strong correlations stated in Case-1 waters.}, \BBP can be viewed as a proxy for suspended particulate matter concentrations \cite{neukermans2012situ, Boss2009IOPsSPM}. 
The Copernicus Marine Environment Monitoring Service (CMEMS) makes such products, including multi-sensor products \cite{datacmems}, freely accessible online via its platform\cite{Cmems}. These operational sea surface parameter products can have large sampling gaps, typically ranging between 30\% and 70\% of missing data for a region such as the Mediterranean Sea, due to the impact of cloud cover on satellite-derived measurements \cite{ioannidis2018intra}.

A variety of algorithms have been proposed to deliver gap-free datasets, using among others low-rank matrix completion methods\cite{candes_2010,nguyen_2019}, geostatistical methods like spatio-temporal Kriging\cite{OLIVER_2014}, Optimal Interpolation (OI)\cite{HOYER_2007}, DInEOF\cite{Beckers_2006}, and eDInEOF\cite{alvera_2009}. Recently, neural mapping schemes have emerged as an attractive solution to address sampling gaps in ocean remote sensing data sets \cite{Beauchamp2022, Barth2022, Wang2022, YUAN2020_DLRemoteSensing, Dorffer_IGARSS2024}. They often suggest potential significant improvement in the reconstruction of gap-free products, especially regarding higher-resolution patterns.
Besides differences in the considered neural architectures, such as U-Nets \cite{UNet2015} and Transformers \cite{Transformer2017}, these approaches may involve various learning strategies: training schemes using simulated OSSE (Observing System Simulation Experiment) datasets versus training schemes using only gappy satellite-derived datasets. A shortcoming of the former is the underlying assumption that simulated OSSE datasets do provide account for the variabilities of real observation data, which does not hold for ocean colour products.



In this article, we explore state-of-the-art neural mapping methods, especially 4DVarNet schemes \cite{4DVarNetFablet2021Learning}, to deliver gap-free sea surface \BBP fields from multi-sensor satellite observations. We used the Mediterranean Sea as a case study with the associated Multi-Sensor sea surface CMEMS ocean colour product \cite{cmems-mediterranean}.
Our main contributions are as follows:
\begin{itemize}
    \item We introduce a patch-based resampling approach to train neural mapping schemes directly from observation-only gappy datasets, {\em i.e.} thereby eliminating the need for real or simulated gap-free reference datasets;
    \item Our benchmarking experiments demonstrate that our proposed training framework and the chosen neural mapping 4DVarNet significantly enhances the reconstruction of sea surface \BBP fields.

    \item We conducted extensive experiments with various satellite sensors (MODIS, VIIRS, SeaWiFS, and OLCI) to evaluate their contributions to data quality. Our analysis confirmed that using all available satellite sensors improves the overall performance. The VIIRS-JPSS1 sensor was found to be particularly crucial thanks to its large swath that covers a larger area.

\end{itemize}

The paper is organized as follows: \Cref{sec:case_study} present the case study, \ie the data, domain, and variables of interest. \Cref{sec:Interp_methods} presents the different interpolation methods that will be tested. \Cref{sec:learn_strat} then present the different data sampling strategies used for learning. \Cref{sec:expe}  reports on a performance analysis of 4DVarNet algorithm considering the different learning/testing setup. Results are compared to state-of-the-art approaches e/DInEOF and direct inversion interpolation.  This section also proposes to evaluate the interpolation improvement reached using different combinations of satellite observations.
Finally, \Cref{sec:conclusion}  synthesizes our main findings and discusses future work.


\section{Case Study}
\label{sec:case_study}
\subsection{Product and Variables of interest}
We used the product \cite{datacmems} provided by the CMEMS that consists of a collection of daily, Multi-Sensor (MODIS-aqua, VIIRS-JPSS1, VIIRS-SNPP, and OLCI-sentinel3a and OLCI-sentinel3b) satellite images acquired from September 1997 to date. 
This dataset consists of daily "level-3" (L3) data, for which images have been regridded, pre-processed (atmospheric corrections were applied), daily merged. It includes backscattering coefficients \BBP, processed from the raw multispectral reflectance data.
The resulting fields involve large missing data rates, typically from 10\% to 80\% depending on the considered space-time location in the case-study region.

\subsection{Data splitting}

The product covers the entire Mediterranean Sea, from \SI{30}{\degree}N to \SI{46}{\degree}N and from \SI{6}{\degree}W to \SI{36.5}{\degree}E, with a \SI{1}{\kilo\meter}$\times$\SI{1}{\kilo\meter} resolution, leading to a grid size of $1580 \times 3308$ pixels. We define a $240\times240$ focus region that extends from  \SI{41}{\degree}N to \SI{43.5}{\degree}N and  \SI{3}{\degree}E to \SI{6}{\degree}E, \ie along the French coast. 

\begin{figure*}[ht]
        \centering
        \includegraphics[width=\linewidth]{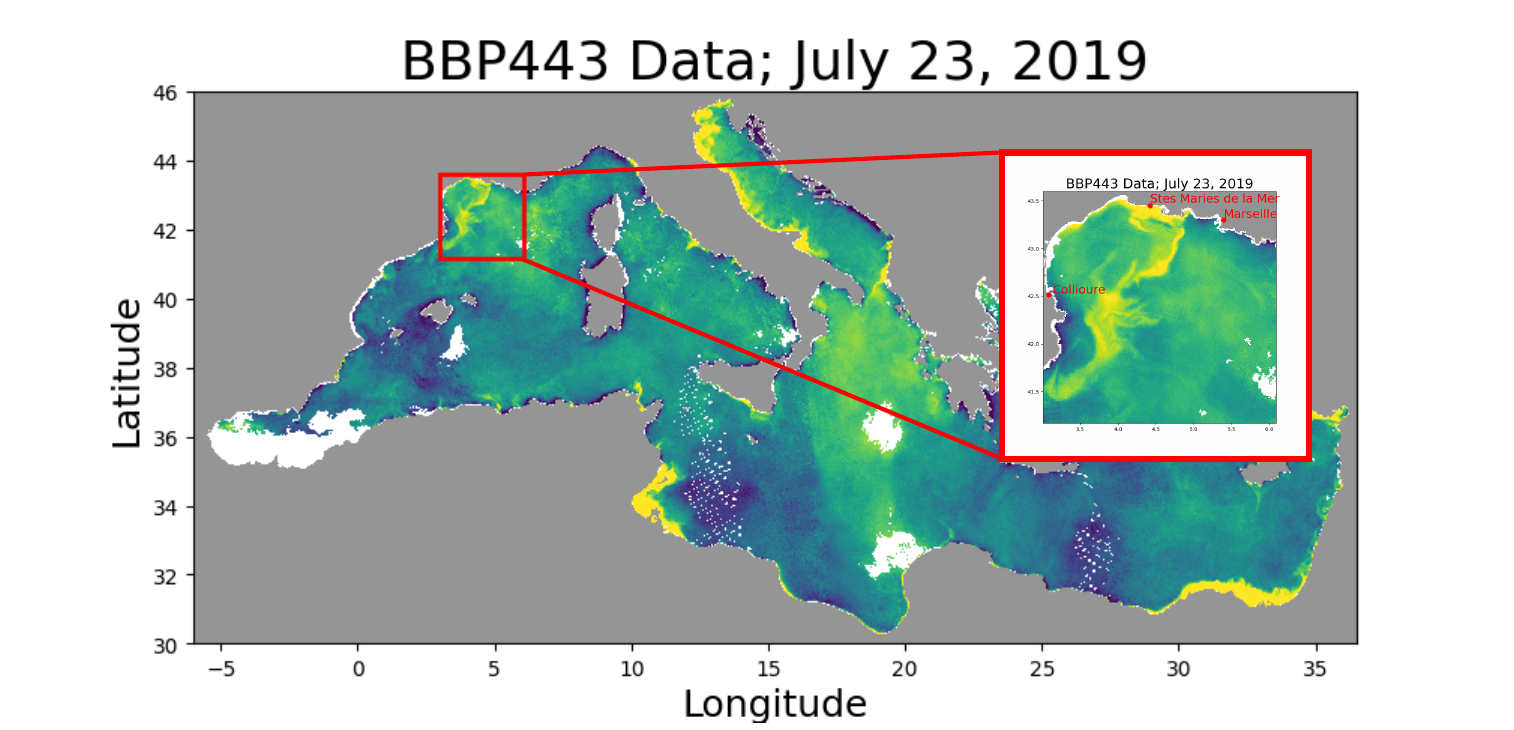}
        \caption{Selected area of interest used for learning and evaluation tasks. Example of \BBP data, in common log values, from July 23th, 2019. Grey pixels are the continental mask, while white ones correspond to missing values.}
        \label{figArea}
    \end{figure*}

\begin{figure}[ht]
\centering
\includegraphics[width=.48\textwidth]{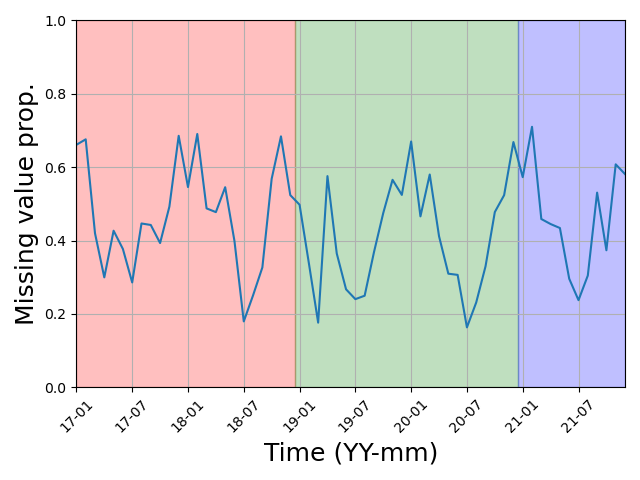}
\caption{Evolution of the monthly mean missing value proportion in the $240\times 240$ area from Jan. 2017 to Dec. 2021 and data splitting schemes used for learning-based approaches. Red, blue and green time windows correspond respectively to the training, validation and testing periods.}
\label{TimeSplit}
\end{figure}

\Cref{figArea} shows an example of log10 \BBP concentration for both the entire area covered by the CMEMS product and the reduced area of interest. The image used is from July 23, 2019, for which missing values account for around $10\%$ of the area, a really small level of incompleteness when compared to the mean missing value proportion of $45\%$ over the time period considered (Jan. 2017 to Dec. 2021). Evolution of the monthly mean missing value proportion over the restricted area is shown in \cref{TimeSplit}. The seasonality of the cloud coverage is clearly visible, with more clouds during the winter than the summer periods. \Cref{TimeSplit} also shows the data splitting used in this study to distinguish the train/test/validation periods.


\section{Interpolation methods}
\label{sec:Interp_methods}
We tested three interpolation methods: Data Interpolating Empirical Orthogonal Functions--DInEOF--\cite{Beckers_2006}, a popular approach for the operational production of L4 satellite images\cite{Volpe_2018}, its enhanced version\cite{alvera_2009} that encompasses a temporal filtering step to force the temporal correlation into the reconstructed field, a UNet-based neural mapping scheme \cite{UNet2015}, and the NN-based Variational Data Assimilation algorithm (4DVarNet)\cite{4DVarNetFablet2021Learning} that has recently shown state-of-the-art interpolation performance for satellite-based Sea Surface Temperature (SST) and Sea Surface Height (SSH) reconstructions\cite{febvre2023training}.

Here, we first briefly introduce the first two approaches, then we provide more details about the 4DVarNet method.

\subsection{DInEOF}

Considering $x_1,\dots,x_n$ a collection of incomplete images focused on the same area with $m$ pixels, then one can build $X\in\R^{m\times n}$ the sparse data matrix whose columns correspond to the vectorized images.
Data Interpolating Empirical Orthogonal Functions, or DInEOF\cite{HOYER_2007}, is an iterative matrix completion approach that consists in:
\begin{enumerate}
    \item Computing $\tilde{X}$ a low-rank approximation of $X$ (using truncated Singular Value Decomposition);
    \item Filling missing entries in $X$ by corresponding values in $\tilde{X}$;
    \item Repeating until convergence.
\end{enumerate}
While relatively simple, this approach provides interesting results, especially when the observed physical phenomenon is slowly changing over time, thus ensuring that most of the signal can be efficiently recovered using only few principal components. However, when using this approach, each image is recovered independently and does not benefit from the temporal dependencies within the dataset. As a consequence, totally blind images or highly sparse images are usually not correctly interpolated using DInEOF.
An enhanced version of this method, eDInEOF, has been proposed in\cite{alvera_2009}, consisting of a filtering step applied after the low-rank approximation to ensure a temporal correlation within matrix $\Tilde{X}$ and thus tending to maintain a temporal continuity in the interpolated images. 

\subsection{Neural mapping}

A first and basic way to use Artificial Intelligence to solve a data interpolation problem consists of feeding a neural scheme with sparse images and optimizing the network weights to minimize the reconstruction error, \ie the difference between the output of the neural network and the target data to be recovered. Once the training stage is done, the interpolated field associated to incomplete time series is simply provided by the output of the ANN fed by the time series. We refe

Different neural architectures can be considered. In the latter, we propose testing a UNet\cite{UNet2015} and a simple Convolutional Neural Network.

\subsection{4DVarNet}

4DVarNet is a Neural Network (NN) based version of the traditional data assimilation approach 4DVar\cite{courtier1994strategy}. It mainly consists in solving the problem
\begin{equation}
    \tilde x = \arg\min_x U(x) = \lambda_1\normOmega{x-y}^2+\lambda_2\normFro{x-\phi(x)}^2,
    \label{eq_varcost}
\end{equation}
where \normOmega{} stands for the $l2$ (possibly $l1$) norm computed on the observation domain $\Omega$, $\lambda_1$ and $\lambda_2$ are the variational cost parameters and $\phi$ is the dynamical model that could be a differential equation model, a physical model, or a NN-based model. 
Resolution of \cref{eq_varcost} is performed using yhr iterative gradient descent update
\begin{equation}
    x^{k+1}=x^k+\rho \nabla U,
\end{equation}
where $\nabla U$ is the gradient of the variational cost \cref{eq_varcost}--that is computed using automatic differentiation tools--and $\rho$, the gradient step-size. Instead of manually choosing $\rho$, which might be challenging, one can think about using a NN-based solver such as Long Short Term Memory networks (or $lstm$ \cite{}) that drives the gradient descent by smoothing successive updates.

\begin{figure*}[ht]
        \centering
        \includegraphics[width=\linewidth]{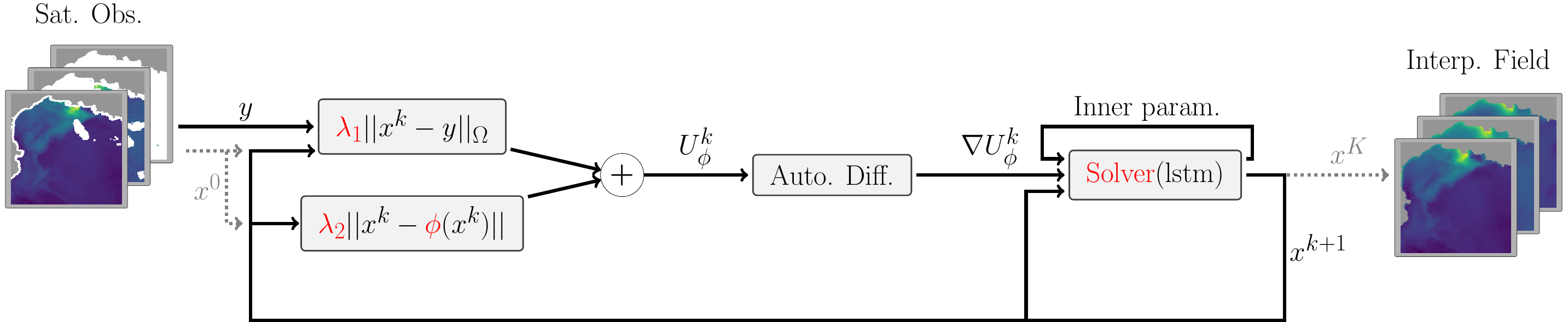}
        \caption{Structure of the 4DVarNet algorithm. Trainable parts \ie the lstm parameters, variational cost parameters ($\lambda_i$), and model parameters are presented in \color{red}red.}
        \label{fig_4dvarStruct}
    \end{figure*}

Considering all these different parts, 4DVarNet algorithm consists in an end-to-end architecture that is presented in \cref{fig_4dvarStruct}.
Interestingly, if one considers using a NN for model $\phi$, then its parameters can be learned simultaneously with variational cost weights and $lstm$ inner-parameters so as to provide the best interpolation. Following the learning scheme presented in \cref{fig_4dvarlearn}, the parameters to be updated with back-propagation are $\lambda_1$, $\lambda_2$, $\phi$ and $lstm$, \ie those appearing in red in \cref{fig_4dvarStruct}.

\section{Learning Strategies}
\label{sec:learn_strat}
\begin{figure}[ht]
        \centering
        \includegraphics[width=\linewidth]{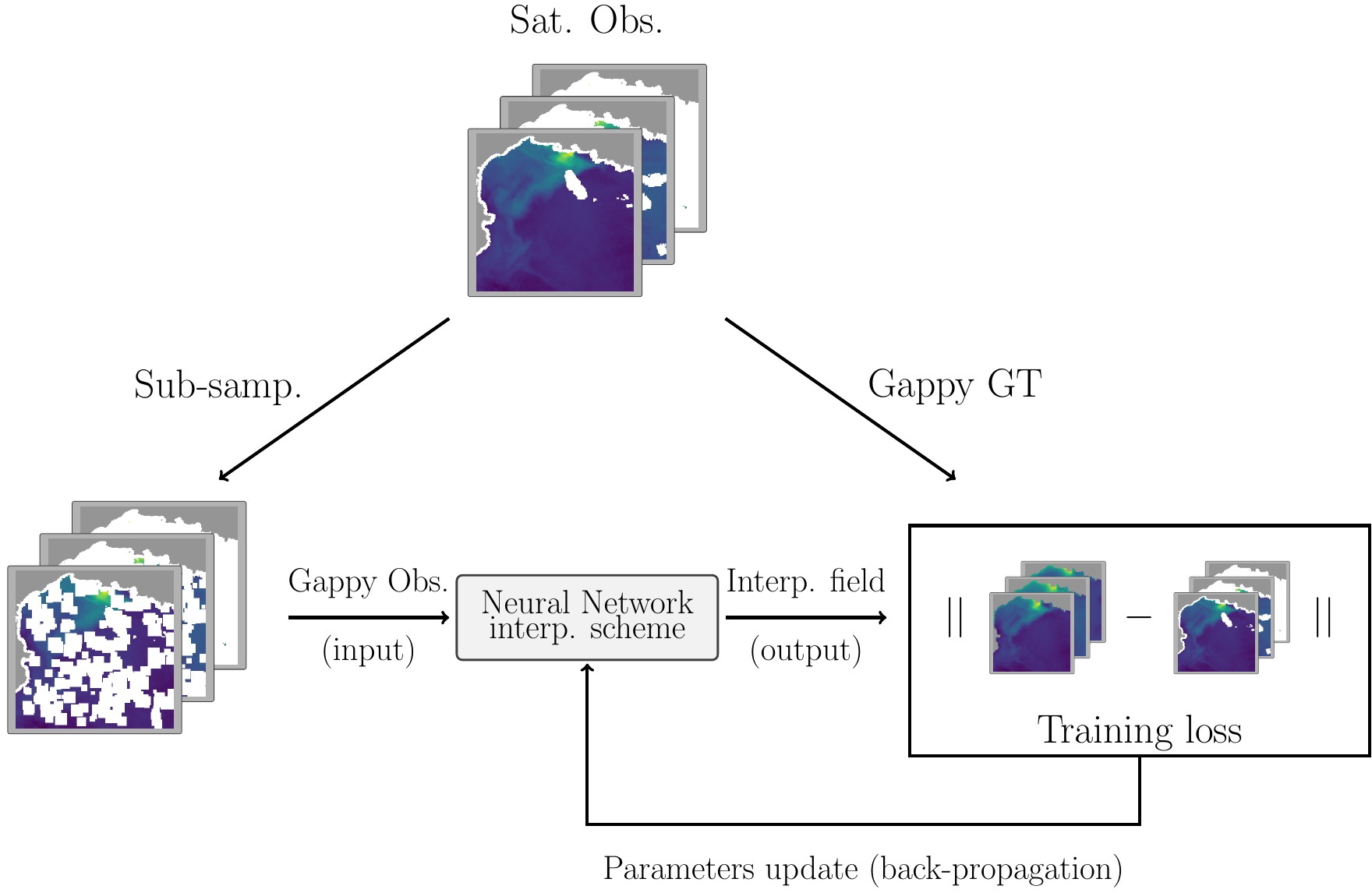}
        \caption{Proposed learning strategy scheme for Neural-Network based approaches.}
        \label{fig_4dvarlearn}
    \end{figure}

Neural Network based approaches require observations and target data to be fitted to during the training phase. Depending on the application, those targets can be \eg known Ground Truth--hardly available for satellite multispectral imaging--or reference model outputs such as in \cite{febvre2023training}. Here, we aim to train NN-based mapping algorithms using only available gappy satellite images. We cannot feed the training algorithms with the same data as input and target without any additional assumptions as the NN-based approaches will tend to learn the identity. We then propose to use the gappy images to artificially generate even more gappy observations. The training data then consist in considering the gappy satellite images as the target and the sub-sampled version of these images as the input data. As such, the training loss is evaluated over all available pixels from the satellite images. An overview of this learning strategy is presented in \cref{fig_4dvarlearn}.

One could think that a pure random sub-sampling of the satellite images--\ie randomly removing pixels from the images--is an easy way to build a training dataset. However, such a sub-sampling scheme does not provide an efficient learning dataset as it defines a too simple interpolation problem and thus does not constrain the NN to extract general patterns. We then 
propose two other resampling strategies: a first one based on real satellite masks; and another one based on the random removal of square patches.

\subsection{Real satellite-based observation patterns}

The data product considered \cite{datacmems} is a Multi-Sensor product, meaning that the images result from the fusion of optical data acquired by different satellite sensors. A sensor mask informing the different sensors contributing to each pixel is provided for each daily image.

\begin{figure}[ht]
        \centering
        \includegraphics[width=\linewidth]{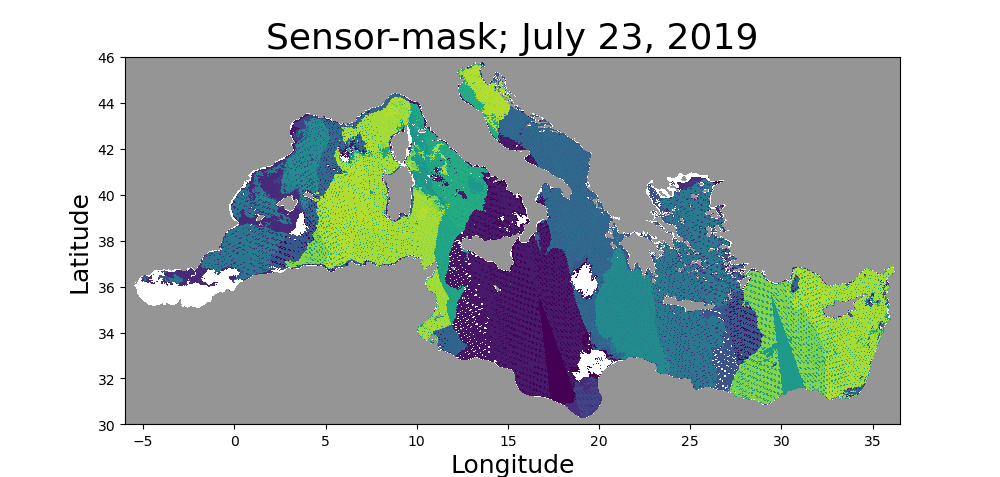}
        \includegraphics[width=\linewidth]{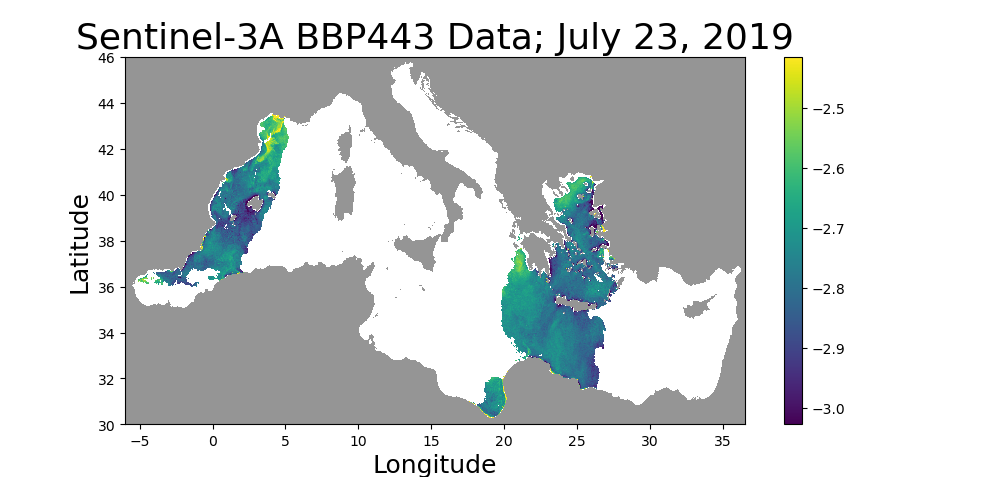}
        \caption{Example of satellite-based observations sampling. (top) Available sensor mask. Color are composed w.r.t. available senseors presented in \cref{sec:case_study}. (bottom) Observations sampled using the Sentinel-3-A mask. The unit of measurement is depicted in log\(_{10}\) scale of $m^{-1}$.}
        \label{sensorMask}
\end{figure}

Our first data sub-sampling strategy exploits these sensor masks to generate gappy patterns associated with different Multi-Sensor configuration as illustrated in \cref{sensorMask}. As an example, we also display the resulting field  when considering only the OLCI-S3A sensor mask is also presented in \cref{sensorMask}. Such a sampling strategy provides realistic observation in the way that produced observations correspond to real observation patterns. As a drawback, it makes the proportion of missing data and spatial covering--that is preferred to be homogeneous--difficult to tune. This strategy also provides only one observation mask per day and does not allow randomized simulations. Moreover, some satellites follow trajectories that do not ensure a dense coverage in time and/or space. For example, Sentinel-3A revisits the same site every two days, which means that for a given pixel an observation is available at most one day out of two, depending on the cloud cover.

\subsection{Randomised patch-based patterns}

Another approach that allows us to better control missing data proportion, spatial sampling, and other factors consists in removing random patches instead of removing random individual pixels. The size and shape of the removed patches can also be controlled. This strategy seems appealing as a trade-off between fully random patterns and realistic ones.

In the latter, we propose generating the observations by removing 50\% of data from the Multi-Sensor images, with original missing value proportions that are below 75\%. In other words, images from the original dataset that have more than 75\% of missing data are kept in full as observations, while half the content of the other images is removed. 
Patch sizes (heights and width) were randomly generated between 5 to 25 pixels, which in our case represents 0.04\% to 1\% of the image.
Overall, this approach allows us to discard enough observation data so that the inputs and targets  depict significant differeneces, while preserving very gappy images in the observation subset.

\begin{figure*}[ht]
    \centering
    \includegraphics[width=.32\textwidth]{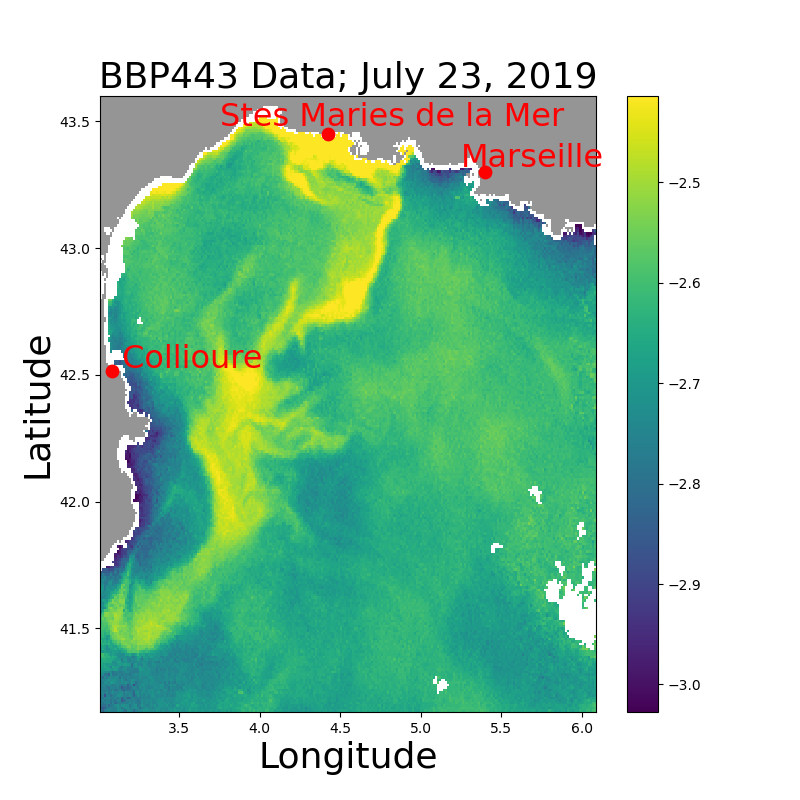}
    \includegraphics[width=.32\textwidth]{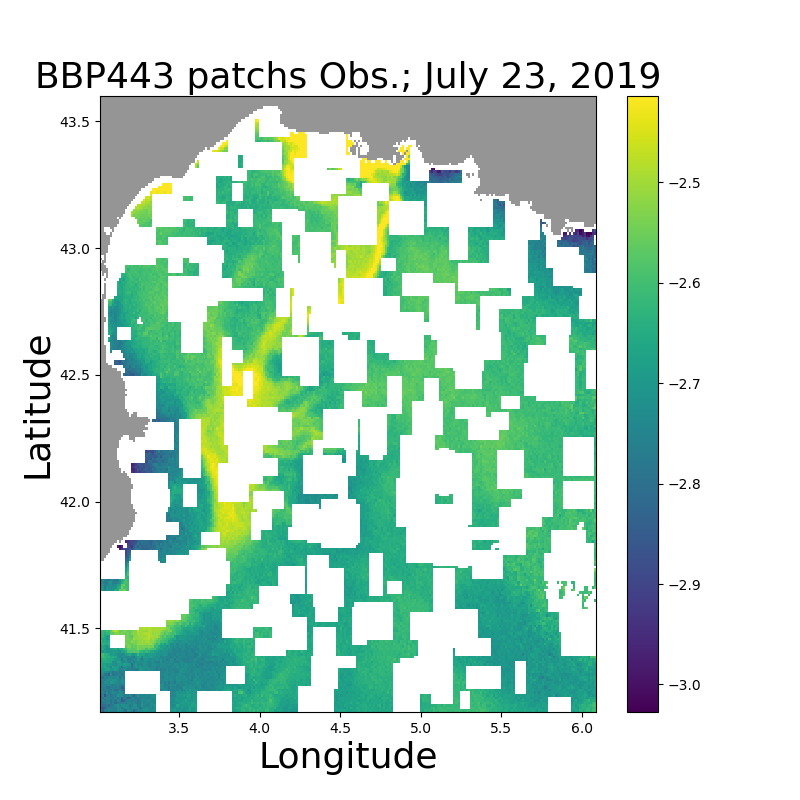}
    \includegraphics[width=.32\textwidth]{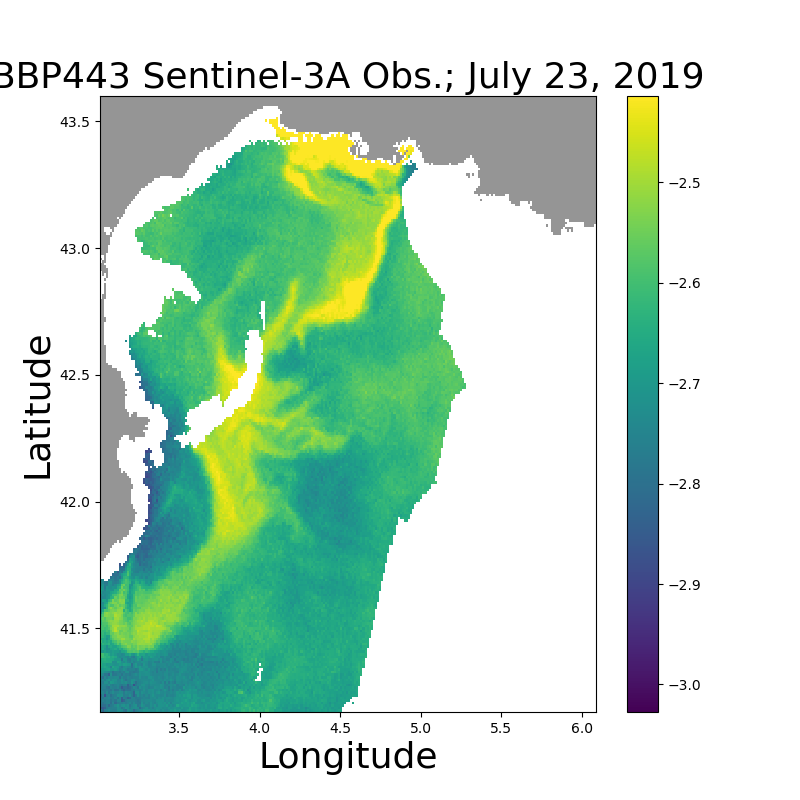}
    \caption{Example of observation masks to be considered in the training phase. (left) Gappy ground-truth, (middle) Sentinel-3A based obs. and (right) random patch obs. The unit of measurement is depicted in log\(_{10}\) scale of $m^{-1}$.}
    \label{fig_obs_masks}
\end{figure*}

For illustration purpose, \Cref{fig_obs_masks} shows one example of the two resampling approaches.


Our code and data are open source and available online at \href{https://github.com/CIA-Oceanix/4dvarnet-core/tree/turbidity-fish-predict}{this link}.

\section{Experiments}
\label{sec:expe}

In our numerical experiments, we benchmark the considered interpolation methods presented using different learning and testing configurations in terms of observations patterns. More specifically, we aim to:
\begin{enumerate}
    \item compare the interpolation performance achieved by DInEOF, Direct-Inversion, and 4DVarNet using a common dataset;
    \item analyze the impact of training and testing observation patterns on the performance of NN-based models;
    \item asses a potential enhancement of the interpolation performance  when using multi-sensor datasets.
\end{enumerate}
For evaluation purposes, we focus on the restricted area shown in \cref{figArea}. We compute for the interpolated fields for the 2-year testing time period that ranges from 2019-01-01 to 2020-12-31 (see \cref{TimeSplit}) with the two following metrics. The Root Mean Squared Log Error (RMSLE):

\begin{equation}
    RMSLE = \sqrt{\frac{1}{\#\Omega}\cdot\sum\limits_{i\in\Omega}(\log_{10}(x(i))-\log_{10}(\tilde{x}(i)))^2},
\end{equation}

and the Mean Relative Error (MRE):

\begin{equation}
MRE(\%) = \frac{1}{\#\Omega}\cdot\sum\limits_{i\in\Omega}100\cdot\left\vert\frac{ x(i)-\tilde{x}(i)}{x(i)}\right\vert.
\end{equation}

In both cases, $\Omega$ is the considered spatio-temporal domain on which the error is evaluated, \ie pixels from the Gappy Ground Truth (GT) that are discarded from the observations. $x(i)$ is the GT values at spatio-temporal location $i$ and $\tilde x(i)$ is the associated interpolated value.

\subsection{Benchmarking of the interpolation methods}

\begin{table}
    \centering
    \begin{tabular}{|c|c|c|c|c|c|}
        \hline
        Algo.             & RMSLE & MRE      \\
        \hline
        \hline
        DInEOF            &  0.0758 & 11.7  \\
        eDInEOF           &  0.0745 & 12.11 \\
        CNN               &  0.128  &  21.9 \\
        UNet              &  0.119  &  18.9 \\
        4DVarNet (CNN)    &  0.0501 & 7.3   \\
        4DVarNet (UNet)   &  0.0520 & 7.65  \\
        \hline
    \end{tabular}
    \caption{Comparison of the different interpolation methods using patch-based observations.}
    \label{benchmark}
\end{table}

\begin{figure}[t]
    \centering
    \includegraphics[width=.49\linewidth]{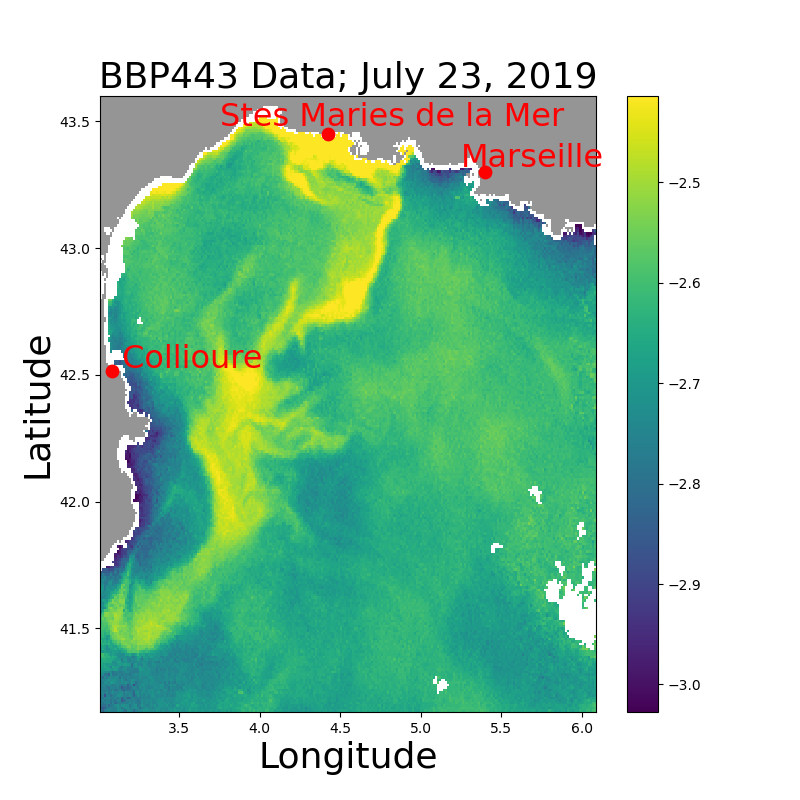}
    \includegraphics[width=.49\linewidth]{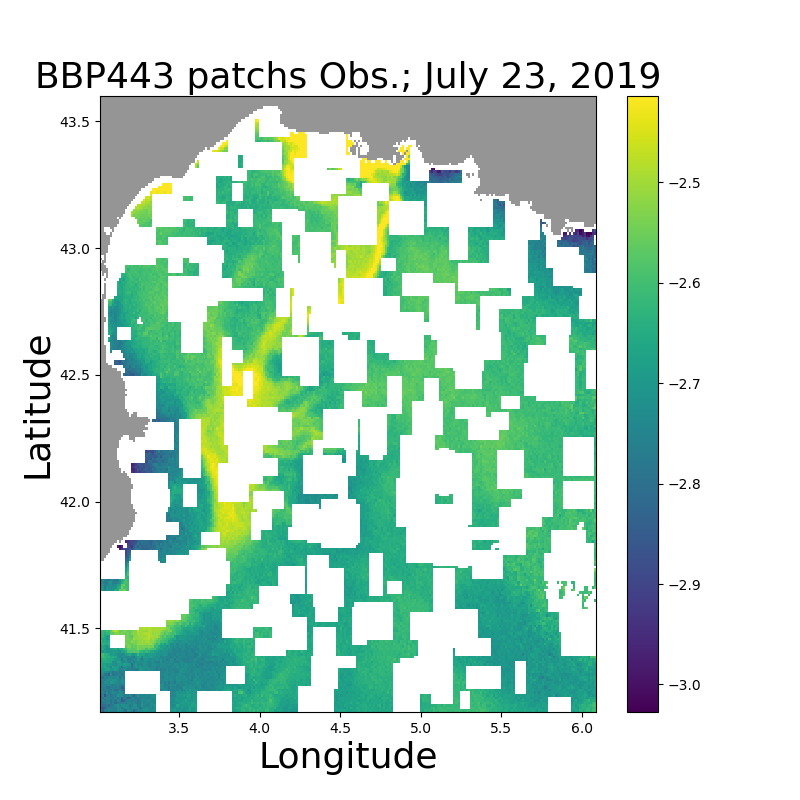}
    
    \includegraphics[width=.49\linewidth]{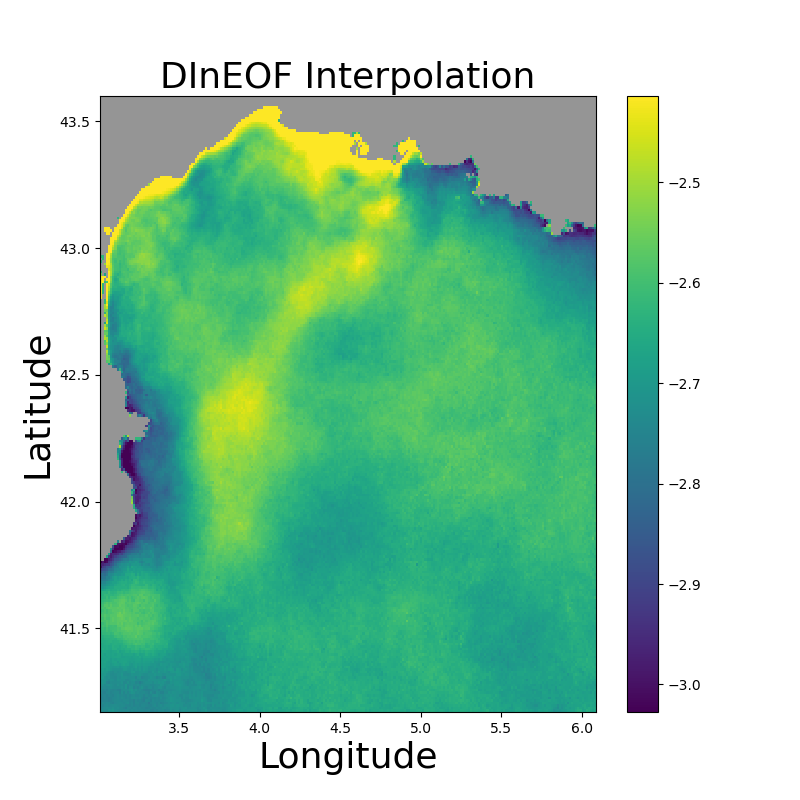}
    \includegraphics[width=.49\linewidth]{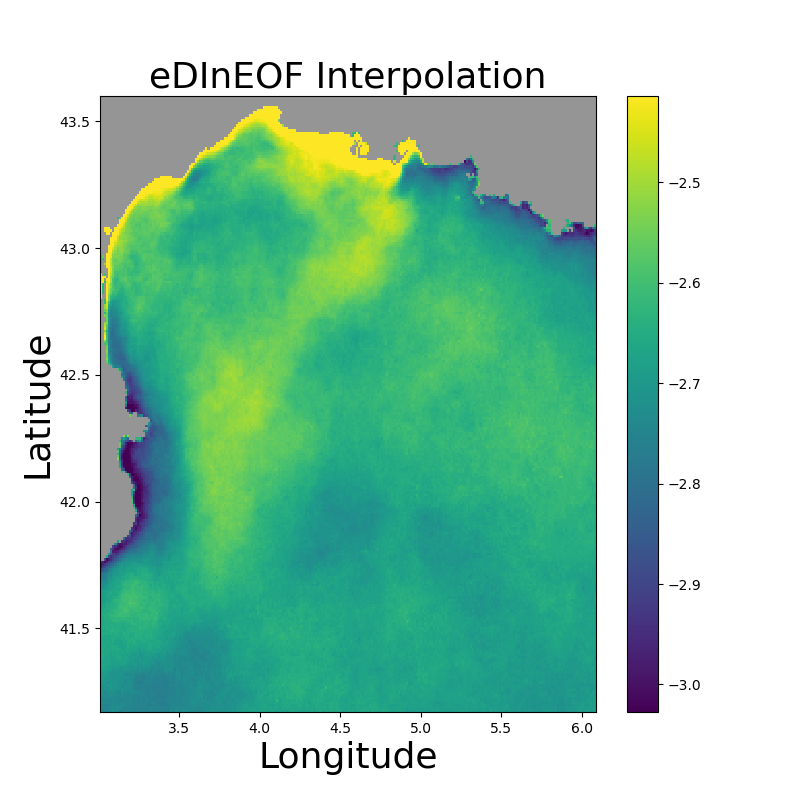}
    
    \includegraphics[width=.49\linewidth]{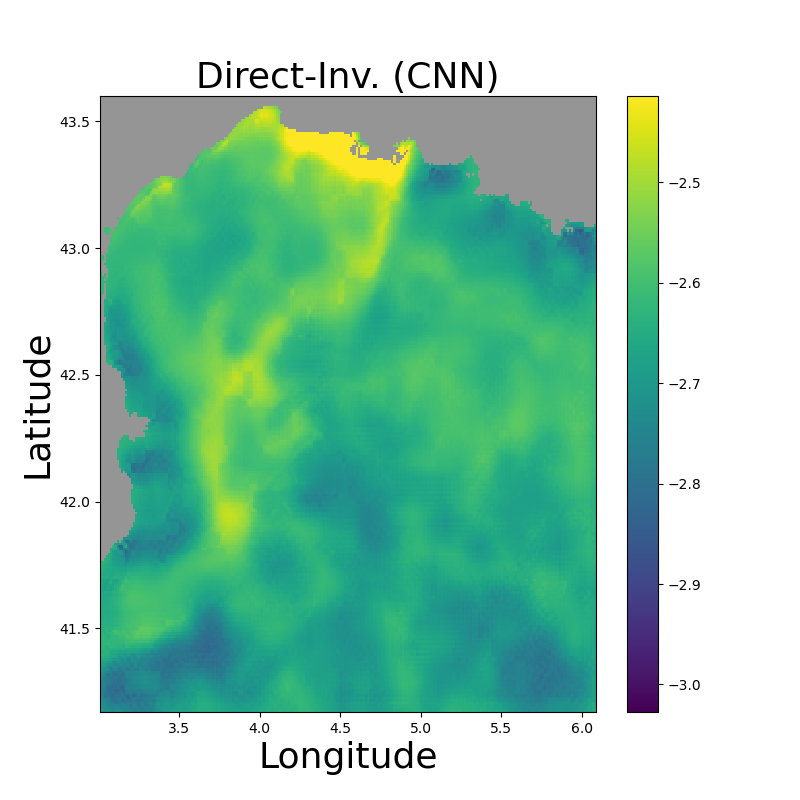}
    \includegraphics[width=.49\linewidth]{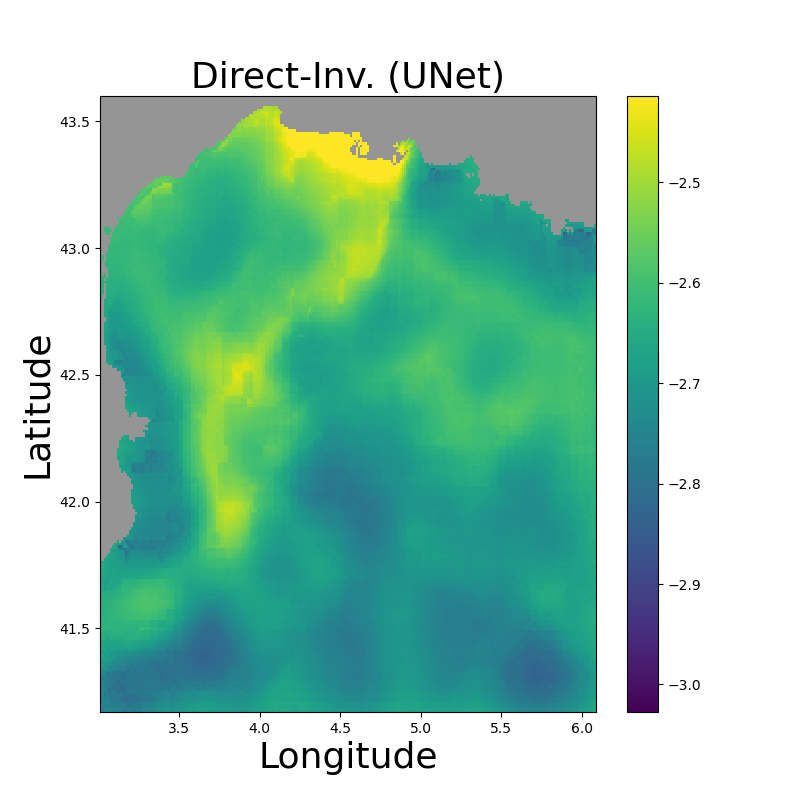}
    
    \includegraphics[width=.49\linewidth]{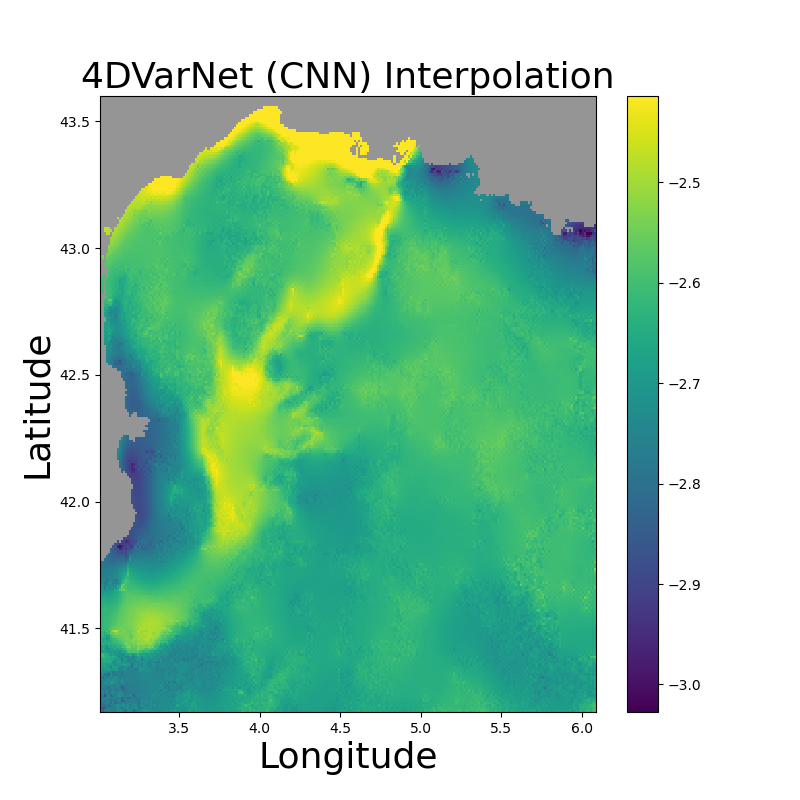}
    \includegraphics[width=.49\linewidth]{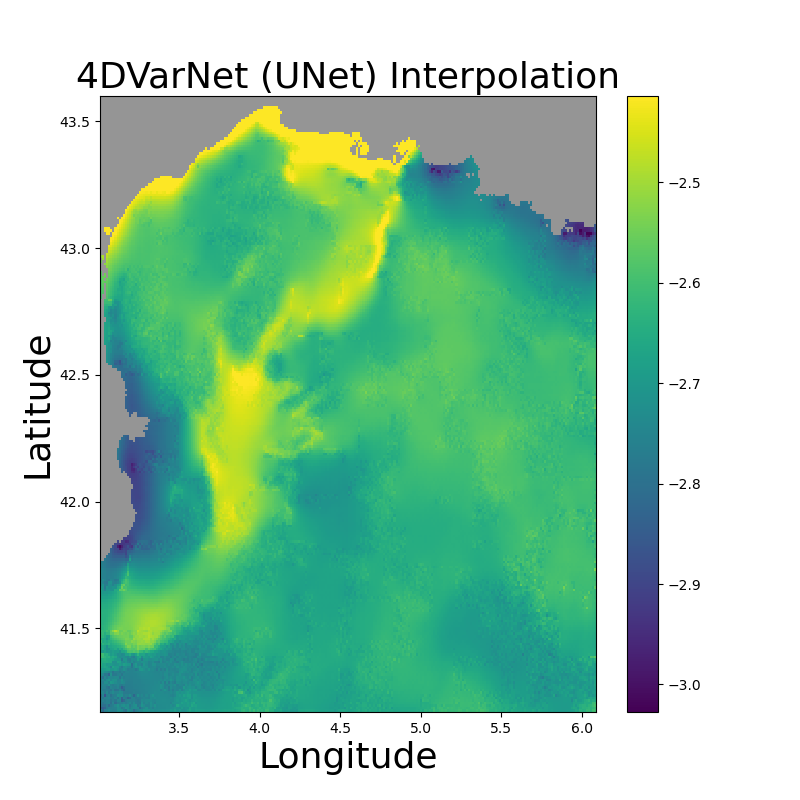}
    \caption{Interpolation examples using different interpolation methods. (top-left) All available observation (target), (top-right) Considered input obs. (2nd line - left) DInEOF interpolation (2nd line - right) eDInEOF interpolation. (3rd line - left) Direct-inversion with CNN. (3nd line - right) Direct-inversion with UNet. (bottom - left) 4DVarNet interpolation with CNN model. (bottom - right) 4DVarNet interpolation with UNet model. The unit of measurement is depicted in log\(_{10}\) scale of $m^{-1}$.}
    \label{Fig:benchmark}
\end{figure}

To compare e/DInEOF, NN-based interpolation, and 4DVarNet, we propose to consider the dataset composed of the multi-satellite images as the Gappy Ground Truth used for the learning step and testing steps, and observations obtained with the random patch-removing strategy for the training and testing steps. Two NN were considered for both the Direct-Inversion and 4DVarNet, a UNet, and a CNN. In both cases networks were trained up to 100 epochs and patch-based observations were randomly generated online during the training stage.

To be fair with non-learning based approaches, especially with e/DInEOF, the same data were used, \ie the gappy GT from the learning stage and the observations from the testing step. In that way, the calibration of the EOF was performed using both training and testing data.

For each algorithm, the interpolation scores were computed over the randomly removed patches of the testing period and were shown in \cref{benchmark}. One can see that 4DVarNet provides a large improvement in performance when compared to other approaches tested, with a 34\% gain in terms of RMSLE, when compare to e/DInEOF interpolation. \Cref{Fig:benchmark} shows a reconstruction example for each approach. One can easily see that 4DVarNet approaches better recover small structures and provides more detailed reconstructions. The smooth aspect of DInEOF and eDInEOF typically comes from the underlying low-rank reconstruction strategy, while reconstructions from neural Direct-Inversion approaches seem to suffer from numerous artifacts.
While 4DVarNets also rely on convolutional architectures, they do not exhibit similar artifacts. This can be explained by the more complex structure of 4DVarNet that consists of an iterative scheme that allows relaxing the "model-fidelity" constraint.

As the 4DVarNet scheme using a CNN as model $\phi$ provides the best performance in these first experiments, we propose to consider only this method for the following tests comparing the impact of the different observation patterns used for the training/testing stages.

\subsection{Impact of the learning setup}

\begin{table}
    \centering
    \begin{tabular}{|c|c|c|c|c|}
        \hline
        Algo.             & Train pattern  & Test Pattern       & RMSLE & MRE      \\
        \hline
        \hline
        4DVarNet (CNN)    &  Patch-based   &                     &  0.0501     &  7.3      \\
        4DVarNet (CNN)    &  Random-based  &   Patch-based       &  0.1479     &  20.7      \\
        4DVarNet (CNN)    &  Sensor-based  &                     &  0.1167     &  21.5      \\
        \hline
        \hline
        4DVarNet (CNN)    &  Patch-based   &                     &  0.0594     &  8.1      \\
        4DVarNet (CNN)    &  Random-based  &   Random-based      &  0.0271     &  4.0      \\
        4DVarNet (CNN)    &  Sensor-based  &                     &  0.1035     &  18.2      \\
        \hline
        \hline
        4DVarNet (CNN)    &  Patch-based   &                     &  0.1669     &  28.6      \\
        4DVarNet (CNN)    &  Random-based  &   Sensor-based      &  0.2361     &  42.8      \\
        4DVarNet (CNN)    &  Sensor-based  &                     &  0.1373     &  25.8      \\
        \hline
    \end{tabular}
    \caption{Root Mean Square Log Error (RMSLE) and Relative Error (RE) reached by the 4DVarNet algorithm, considering different observation setup for training and testing steps.}
    \label{patterns}
\end{table}

\begin{figure}
    \centering
    \includegraphics[width=.49\linewidth]{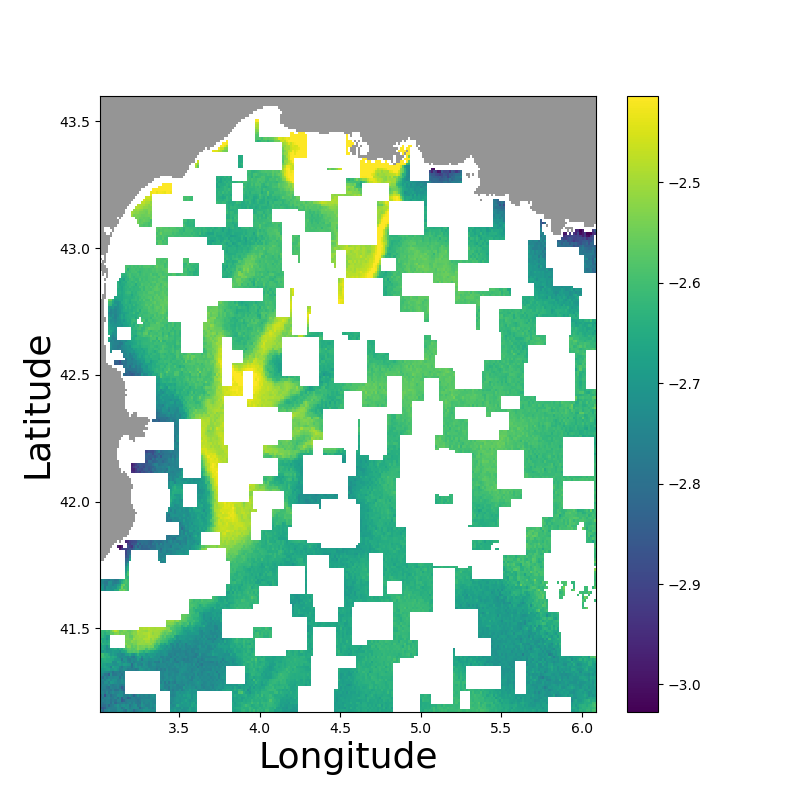}
    \includegraphics[width=.49\linewidth]{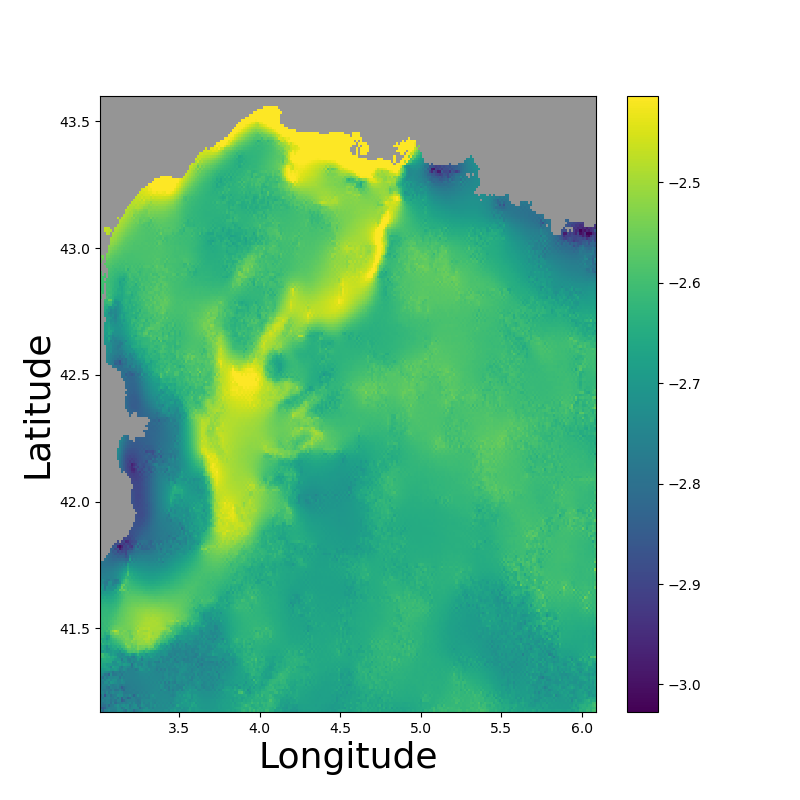}
    
    \includegraphics[width=.49\linewidth]{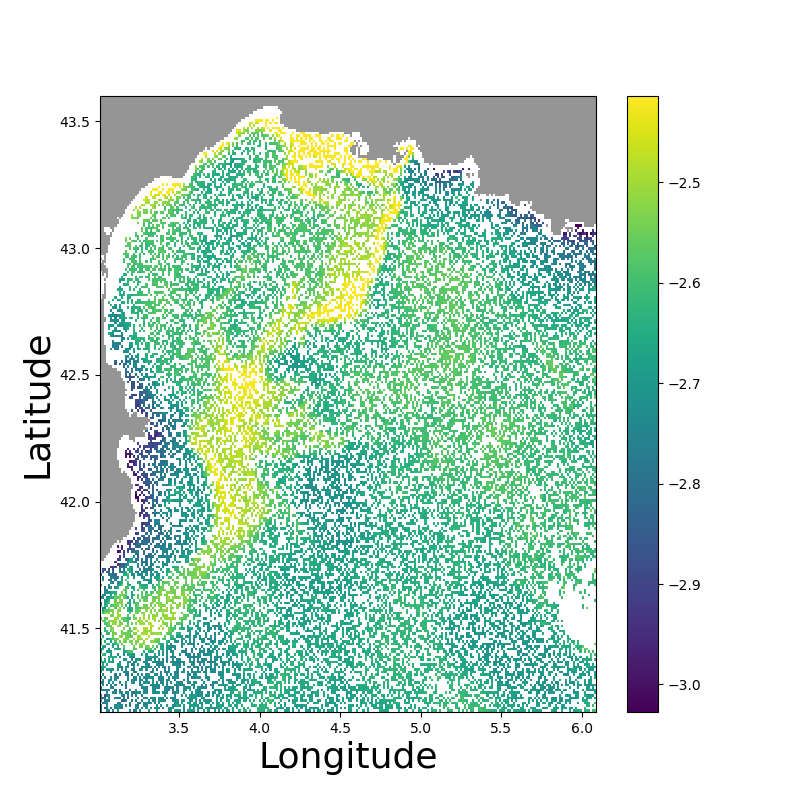}
    \includegraphics[width=.49\linewidth]{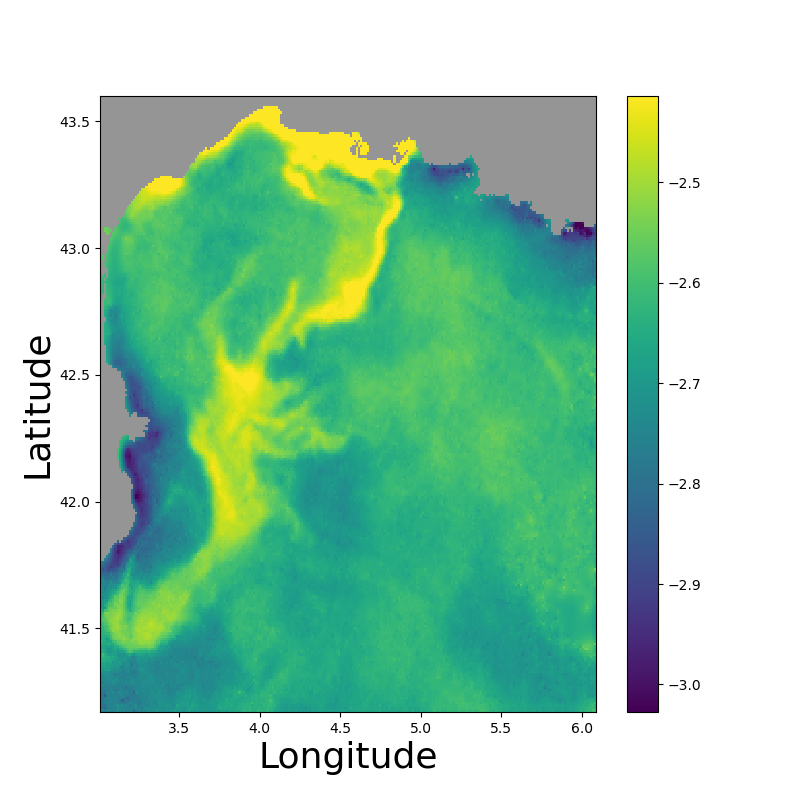}
    
    \includegraphics[width=.49\linewidth]{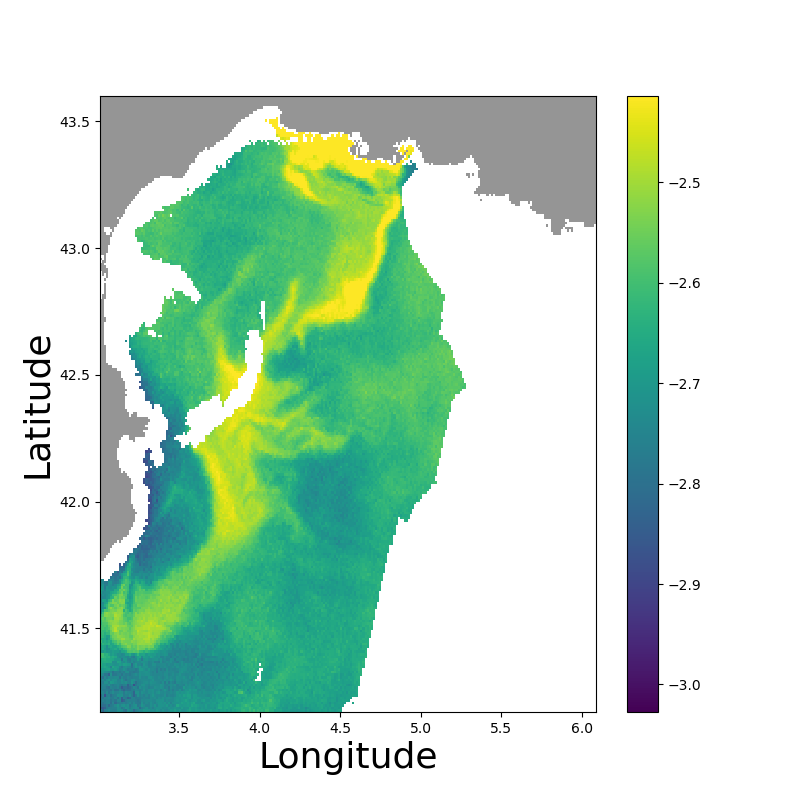}
    \includegraphics[width=.49\linewidth]{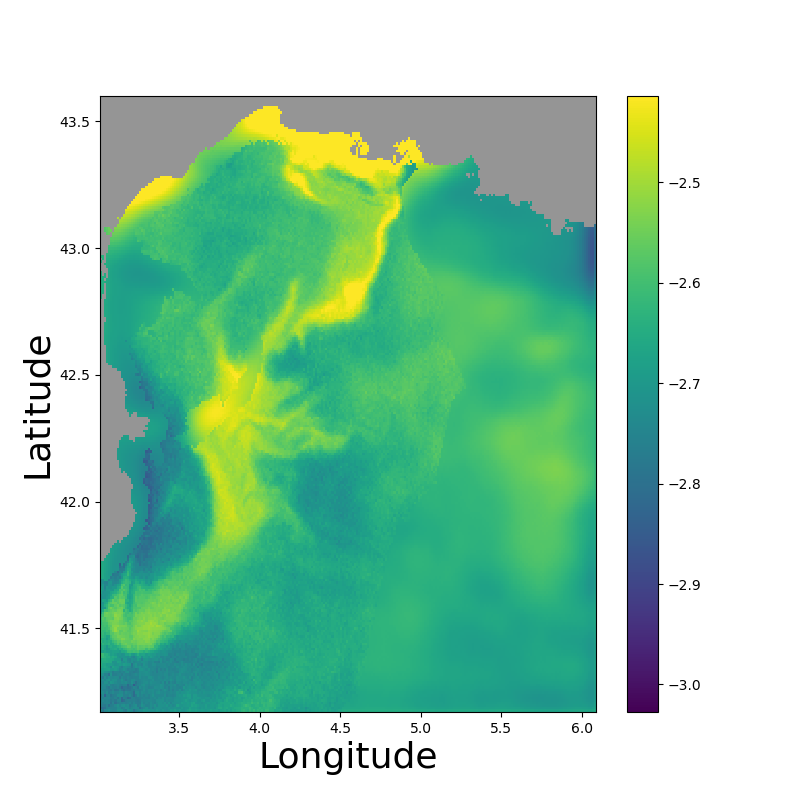}
    \caption{Examples of reconstructions obtained with 4DVarNet for different observation patterns. (left) Considered input observations, (right) associated 4DVarNet reconstructions. The unit of measurement is depicted in log\(_{10}\) scale of $m^{-1}$.}
    \label{Fig:patterns}
\end{figure}

Here, we compare the efficiency of 4DVarNets trained with the different observation patterns proposed in \cref{sec:learn_strat}. To that end, we trained the same 4DVarNet architecture \ie 4DVarNet with the same CNN-based model, using input observations generated with i) the patch-based strategy, ii) a purely random pixel-level strategy where 50\% of data were randomly removed, and iii) the S3A-sensor based strategy. Each trained 4DVarNet was then used for the interpolation of all observation patterns. Interpolation scores were computed on the removed observations, \ie i) on the removed patches for patch-based observations, ii) on the randomly removed pixels for purely random-based observations, and iii) on the non S3A-provided data for satellite-based observations.

We report in \cref{patterns} a summary table of the  interpolation scores reached for all models on all observation strategies. As the scores are computed on different data, depending on the testing observation patterns, one cannot directly compare the performances reached for the different testing observation setups. However, one can compare the performances of the different 4DVarNet models when considering the same testing setup. We point out two interesting results from \cref{patterns}. First, 4DVarNet trained using a given observation pattern always provides the best performance when applied to the interpolation of the same observation pattern. That result shows that during the training phase, 4DVarNets  learn interpolation schemes optimized for the associated training observation pattern. Second, one can see that when it is used for training, the patch-based observation strategy seems to provide an interpolation scheme that generalizes well considering the other two testing observation patterns. the 4DVarNet trained with patch-based observations provides results that are always between the best and worst models, contrary to models learned with purely random or satellite-based observations, that provide poor interpolation scores when used on other setup.

\Cref{Fig:patterns} shows reconstruction examples for all kind of observation patterns using the 4DVarNet model that provides the best RMSLE score for each setup.



\subsection{Impact of the satellite selection}

To assess the improvement due to different sensor contributions that impact the spatio-temporal coverage of the area, we propose a new evaluation framework. First of all, to compare interpolation performance using different input observations, we propose to discard $10\%$ of data from the Gappy Ground Truth, 
ensuring a common evaluation support for each dataset. We then used 4DVarNet algorithm trained with the above patch pattern to interpolate from the observation generated by keeping pixels provided by different sensor combinations, \ie we first interpolate data provided by the different individual sensors, we then moved to data provided by all combinations of two sensors, then by all combinations of three sensors. As the dataset contains data provided by 5 sensors for the considered testing period, the proposed experiment consisted in interpolating 31 datasets and evaluating the reconstruction performances on the common missing data \ie on the $10\%$ removed patches. RMSLE and MRE performances are shown in \cref{tab:SensorSelect}.

Considering the single sensor setup, One can see that interpolations based on the VIIRS-JPSS1 sensor provide the best reconstruction performances, with relative gains of approximately 52\%, 50\%, 19\%, and 7\% compared to OLCI-S3B, OLCI-S3A, MODIS-aqua, and VIIRS-SNPP, respectively.Interestingly, some sensors seem to provide more complementary information than others to the interpolation. Considering the three sensors setup, one can note that each combination of sensors provides approximately the same proportion of missing values. However, interpolation performance is quite different, \eg combination 1+3+4 and 2+3+4 both get 51\% of missing data but respectively raise RMSLE of 0.047 and 0.050, thus showing that the distribution of missing values largely impacts the interpolation results. These results seem consistent with the satellite properties, as sensor 1 (OLCI-S3A) provides finer observations near the coast compared to sensor 2 (MODIS).

\begin{table}[h]
    \centering
    \begin{tabular}{|c|c|c|c|}
        \hline
        Sensor combinaison   & RMSLE & MRE   & MV prop.   \\
        \hline
        \hline
        All-sensors   &   0.04426    &  6.71   & 0.46  \\
        \hline
        \hline
        1 : OLCI-S3A &    0.1372    & 21.82  &  0.84 \\
        2 : MODIS-Aqua       &    0.08516   & 12.77  & 0.67\\
        3 : VIIRS-JPSS1 &\textbf{    0.06891}   & 10.14  & 0.64\\
        4 : VIIRS-SNPP   &    0.07434   & 10.84  & 0.66\\
        5 : OLCI-S3B &    0.1444    & 21.60  & 0.84\\
        \hline
        \hline
        1+2 & 0.06900   &    10.17   &  0.61\\
        1+3 & 0.06158   &    8.99    &  0.59\\
        1+4 & 0.06386   &    9.26    &  0.60\\
        1+5 & 0.09895   &    14.01   &  0.70\\
        2+3 & 0.05905   &    8.62    &  0.57\\
        2+4 & 0.05900   &    8.67    &  0.57\\
        2+5 & 0.07301   &    10.67   &  0.61\\
        3+4 & \textbf{0.05154}   &    7.62   &   0.54\\
        3+5 & 0.06039   &    8.85   &   0.59\\
        4+5 & 0.06441   &    9.31   &   0.60\\
        \hline
        \hline
        1+2+3 & 0.05330 &    7.84   &  0.53  \\
        1+2+4 & 0.05237 &    7.78   &  0.53  \\
        1+2+5 & 0.05868 &    8.54   &  0.55  \\
        1+3+4 & \textbf{0.04756} &    7.14   &  0.51  \\
        1+3+5 & 0.05367 &    7.87   &  0.54  \\
        1+4+5 & 0.05527 &    8.05   &  0.55  \\
        2+3+4 & 0.05045 &    7.46   &  0.51  \\
        2+3+5 & 0.05369 &    7.87   &  0.53  \\
        2+4+5 & 0.05373 &    7.90   &  0.54  \\
        3+4+5 & 0.04807 &    7.16   &  0.51  \\
        \hline
        \hline
        1+2+3+4 & 0.04689  & 7.03   &  0.49 \\
        1+2+3+5 & 0.04865  & 7.22   &  0.50 \\
        1+2+4+5 & 0.04791  & 7.17   &  0.50 \\
        1+3+4+5 & \textbf{0.04471}  & 6.77   &  0.48 \\
        2+3+4+5 & 0.04724  & 7.05   &  0.49 \\
        \hline
    \end{tabular}
    \caption{4DVarNet interpolation performance considering different satellite combinations. Columns correspond to (from left to right): the considered sensor combination, the Root Mean Square Log Error, the Mean Relative Error and the Missing Value proportion.}
    \label{tab:SensorSelect}
\end{table}

\begin{figure}
    \centering
    \includegraphics[width=.49\linewidth]{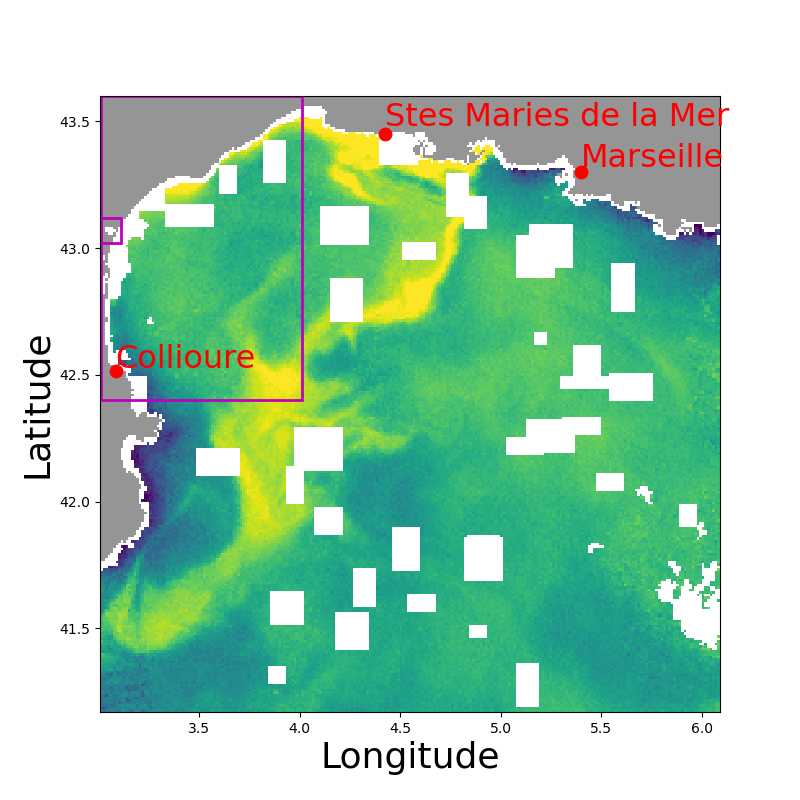}
    \includegraphics[width=.49\linewidth]{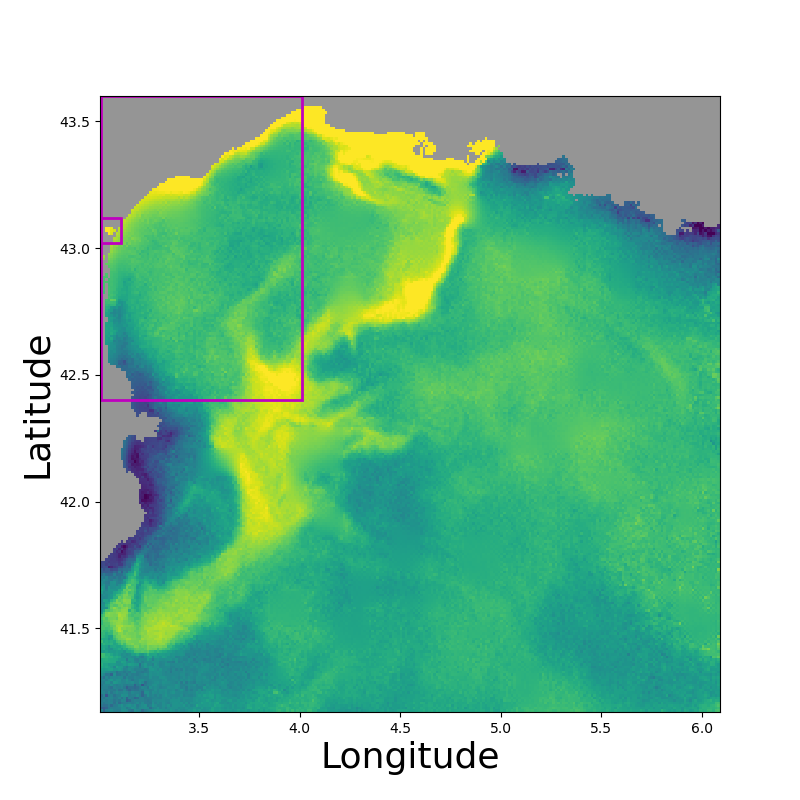}
    
    \includegraphics[width=.49\linewidth]{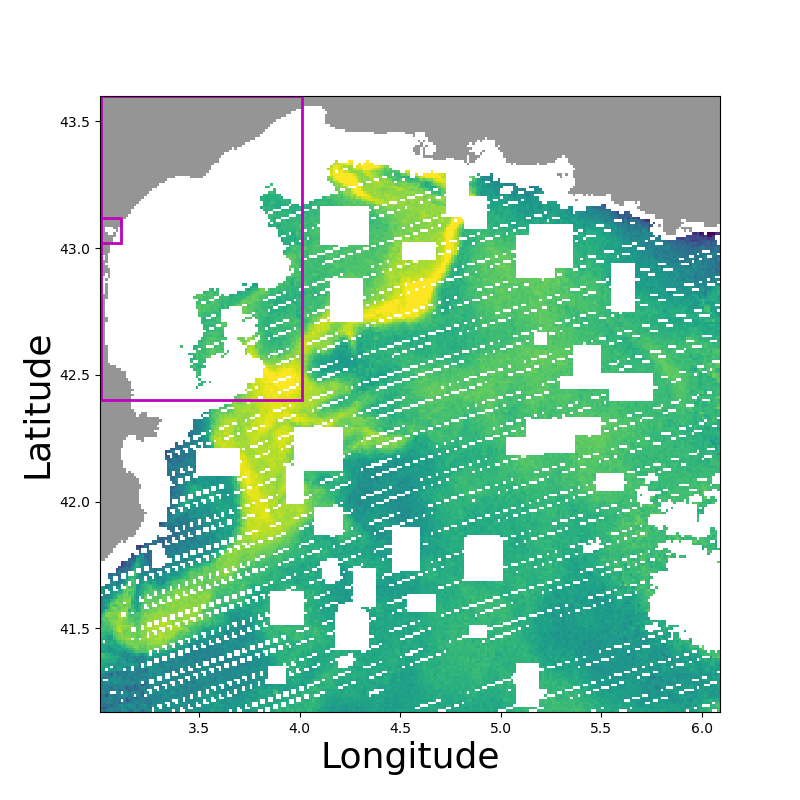}
    \includegraphics[width=.49\linewidth]{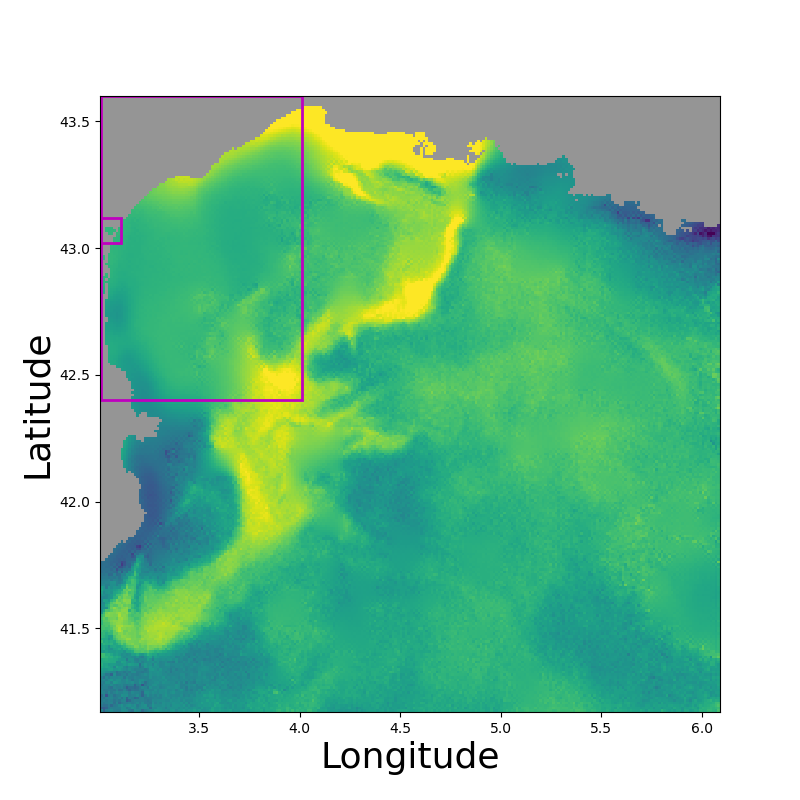}
    
    \includegraphics[width=.49\linewidth]{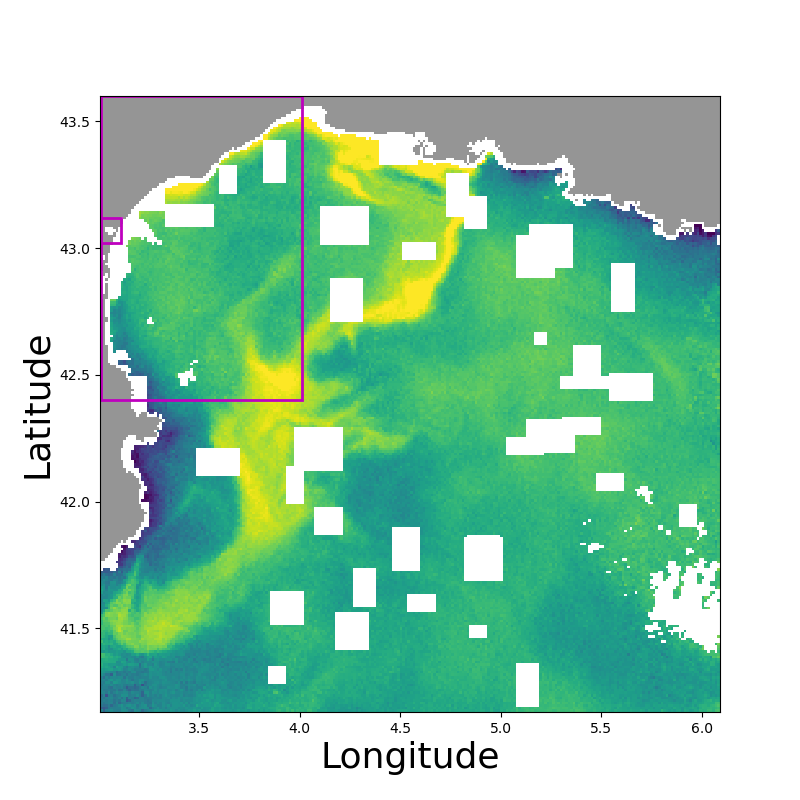}
    \includegraphics[width=.49\linewidth]{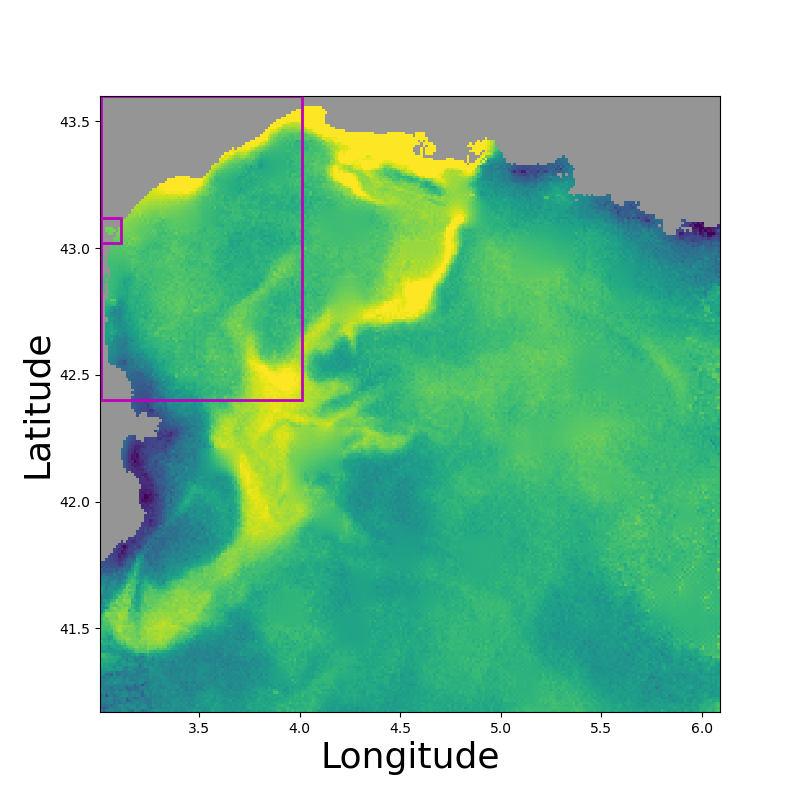}
    
    \includegraphics[width=.49\linewidth]{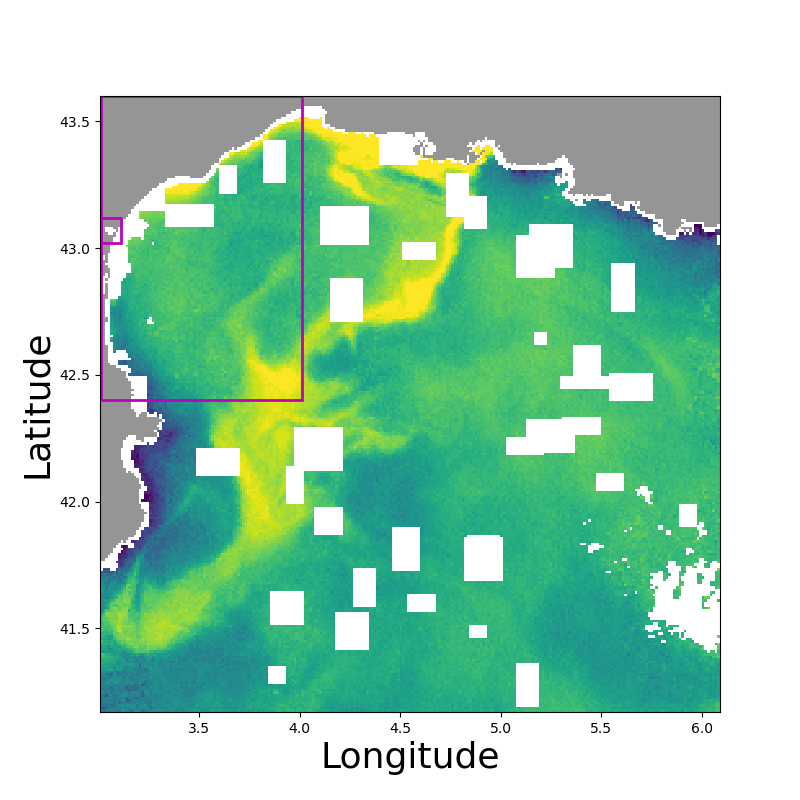}
    \includegraphics[width=.49\linewidth]{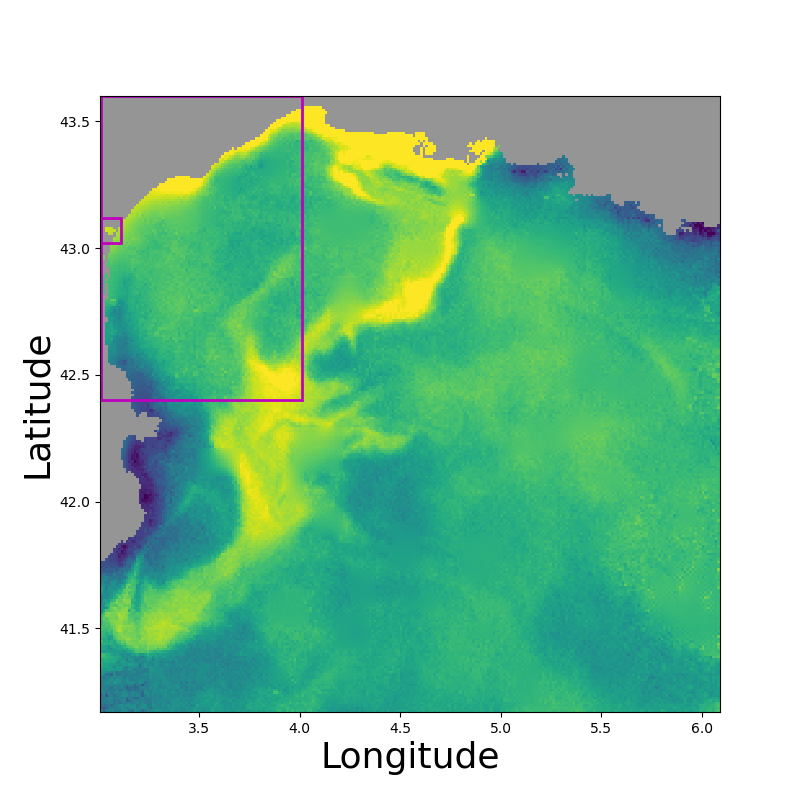}
    
    \includegraphics[width=.49\linewidth]{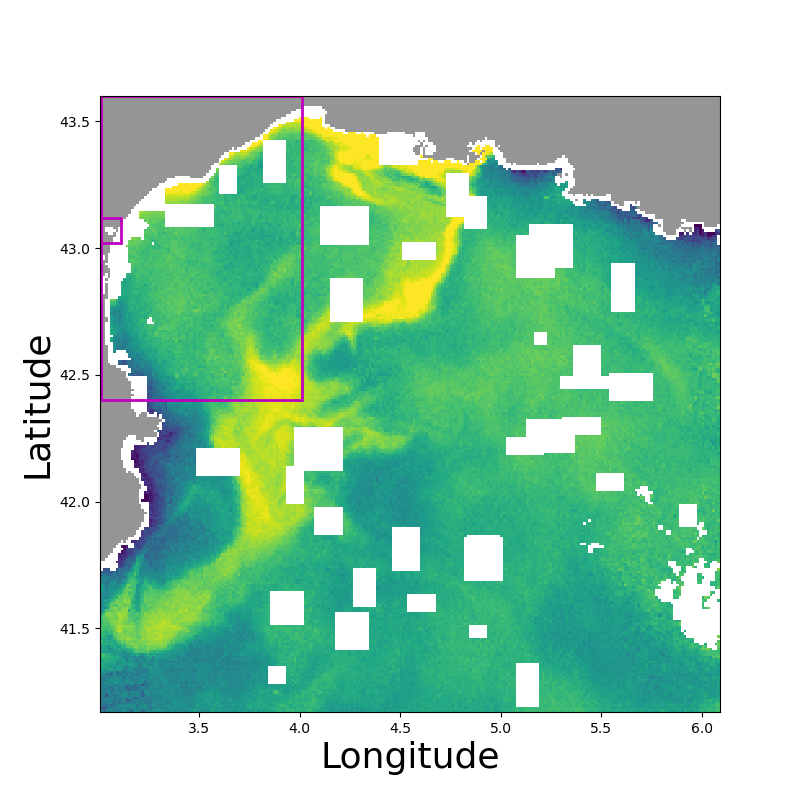}
    \includegraphics[width=.49\linewidth]{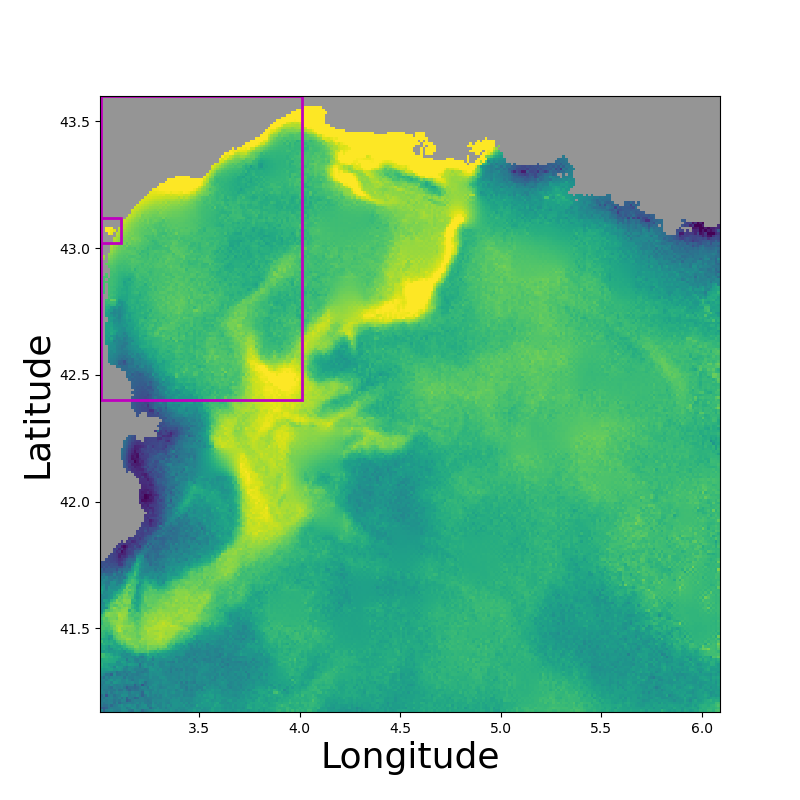}
    \caption{Interpolation examples using different sensors combinations. (left) Observations, (right) 4DVarNet reconstruction. (top) All sensors observation (2nd line) VIIRS-JPSS1 Obs. (3rd line) VIIRS-JPSS1 + VIIRS-SNPP Obs. (4th line) VIIRS-JPSS1 + VIIRS-SNPP +S3A Obs. (bottom) VIIRS-JPSS1 + VIIRS-SNPP + S3A + S3B Obs. The unit of measurement is depicted in log\(_{10}\) scale of $m^{-1}$.  The two boxes emphasize areas with noticeable differences.}
    \label{Fig:sensor_select}
\end{figure}


\Cref{Fig:sensor_select} shows interpolation examples obtained with 4DVarNet algorithm using increasing sensor combinations, \ie by gradually including sensors into the observations. One can see that VIIRS-JPSS1 sensor provides a large coverage of the area but is subject to missing values caused by the acquisition process (striping and bow-tie effects \cite{mikelsons2014destriping}). The addition of VIIRS-SNPP observations leads to a significant improvement for the North-Eastern part of the considered region. The addition of S3A and S3B only brings minor changes, as highlighted for the inland waters along the northeastern sea shore. This is line with the relative improvements brought by the different sensors reported in Table \ref{tab:SensorSelect}.


\section{Conclusion and future work}
\label{sec:conclusion}

In this paper, we demonstrated the effectiveness of training neural networks 4DVarNet schemes directly on gappy data from satellite observations. Through extensive testing across various sampling scenarios, we find that the best-performing approach uses random rectangular patches. For specific application with \BBP reconstructions, our chosen deep neural network, 4DVarNet, outperforms other state-of-the-art methods (e.g.,DInEOF, neural network based Direct Inversion) thanks to its integration of data assimilation within the network.

Based on extensive experiments with different satellite sensors, we have made two key observations: a) Utilizing data from all available satellite sensors yields the best performance thanks to the more available data; b) The VIIRS-JPSS1 satellite sensor appears to be the most crucial, as experiments including VIIRS-JPSS1 data consistently outperform comparable configurations without it. This can be attributed to the particularly large swath of this sensor giving a larger coverage of the earth and so add much more input data in the process.

Although this paper focuses on a restricted area of the Mediterranean Sea, as shown in \Cref{figArea}, it is important to note that, according to \cite{Dorffer_IGARSS2024}, the proposed training framework with gappy groundtruth data and random patch-sampling schemes allows the 4DVarNet architecture to generalize well to other regions and parameters. Especially, while training over the entire Mediterranean can be extensively expensive due to its large area, the study \cite{Dorffer_IGARSS2024} demonstrates a cost-effective strategy. The model trained on the relatively small selected area transfers well to the entire Mediterranean Sea without no fine-tuning step. Similarly, our study supports the potential training of generic interpolation schemes that could apply to a wide range of observation patterns. Future work will likely explore such generalization properties up to the global scale and across ocean colour variables. In this context, the exploration of conditioning variables such as the space-time location, bathymetry as well as other sea surface variables (e.g., sea surface currents, sea surface temperature, sea surface winds,...), could also be highly beneficial. 

Beyond the surface of the ocean, combining the learning-based method of \cite{sauzede2016neuralBBP} to our results open new research avenues to retrieve the full 3D+t picture of \BBP on global scale and fully exploit the potential of the global network of Bio-Argo floats \cite{boss2008observations}.


\bibliographystyle{IEEEtran}
\bibliography{biblio}

\begin{thebibliography}{10}
\providecommand{\url}[1]{#1}
\csname url@samestyle\endcsname
\providecommand{\newblock}{\relax}
\providecommand{\bibinfo}[2]{#2}
\providecommand{\BIBentrySTDinterwordspacing}{\spaceskip=0pt\relax}
\providecommand{\BIBentryALTinterwordstretchfactor}{4}
\providecommand{\BIBentryALTinterwordspacing}{\spaceskip=\fontdimen2\font plus
\BIBentryALTinterwordstretchfactor\fontdimen3\font minus \fontdimen4\font\relax}
\providecommand{\BIBforeignlanguage}[2]{{%
\expandafter\ifx\csname l@#1\endcsname\relax
\typeout{** WARNING: IEEEtran.bst: No hyphenation pattern has been}%
\typeout{** loaded for the language `#1'. Using the pattern for}%
\typeout{** the default language instead.}%
\else
\language=\csname l@#1\endcsname
\fi
#2}}
\providecommand{\BIBdecl}{\relax}
\BIBdecl

\bibitem{mcclain2009decade}
C.~R. McClain, ``A decade of satellite ocean color observations,'' \emph{Annual Review of Marine Science}, vol.~1, no.~1, pp. 19--42, 2009.

\bibitem{dutkiewicz2019ocean}
S.~Dutkiewicz, A.~E. Hickman, O.~Jahn, S.~Henson, C.~Beaulieu, and E.~Monier, ``Ocean colour signature of climate change,'' \emph{Nature communications}, vol.~10, no.~1, p. 578, 2019.

\bibitem{gohin2020satellite}
F.~Gohin, P.~Bry{\`e}re, A.~Lefebvre, P.-G. Sauriau, N.~Savoye, V.~Vantrepotte, Y.~Bozec, T.~Cariou, P.~Conan, S.~Coudray \emph{et~al.}, ``Satellite and in situ monitoring of chl-a, turbidity, and total suspended matter in coastal waters: Experience of the year 2017 along the french coasts,'' \emph{Journal of Marine Science and Engineering}, vol.~8, no.~9, p. 665, 2020.

\bibitem{Bisson2020bbp}
\BIBentryALTinterwordspacing
K.~M. Bisson, E.~Boss, P.~J. Werdell, A.~Ibrahim, and M.~J. Behrenfeld, ``Particulate backscattering in the global ocean: A comparison of independent assessments,'' \emph{Geophysical Research Letters}, vol.~48, no.~2, p. e2020GL090909, 2021, e2020GL090909 2020GL090909. [Online]. Available: \url{https://agupubs.onlinelibrary.wiley.com/doi/abs/10.1029/2020GL090909}
\BIBentrySTDinterwordspacing

\bibitem{mohseni2022ocean}
F.~Mohseni, F.~Saba, S.~M. Mirmazloumi, M.~Amani, M.~Mokhtarzade, S.~Jamali, and S.~Mahdavi, ``Ocean water quality monitoring using remote sensing techniques: A review,'' \emph{Marine environmental research}, vol. 180, p. 105701, 2022.

\bibitem{bricaud1995situ}
A.~Bricaud, C.~Roesler, and J.~R.~V. Zaneveld, ``In situ methods for measuring the inherent optical properties of ocean waters,'' \emph{Limnology and Oceanography}, vol.~40, no.~2, pp. 393--410, 1995.

\bibitem{barnes2003status}
W.~L. Barnes, X.~Xiong, and V.~V. Salomonson, ``Status of terra modis and aqua modis,'' \emph{Advances in Space Research}, vol.~32, no.~11, pp. 2099--2106, 2003.

\bibitem{donlon2012global}
C.~Donlon, B.~Berruti, A.~Buongiorno, M.-H. Ferreira, P.~F{\'e}m{\'e}nias, J.~Frerick, P.~Goryl, U.~Klein, H.~Laur, C.~Mavrocordatos \emph{et~al.}, ``The global monitoring for environment and security (gmes) sentinel-3 mission,'' \emph{Remote sensing of Environment}, vol. 120, pp. 37--57, 2012.

\bibitem{Hieronymi_2023}
\BIBentryALTinterwordspacing
M.~Hieronymi, S.~Bi, D.~Müller, E.~M. Schütt, D.~Behr, C.~Brockmann, C.~Lebreton, F.~Steinmetz, K.~Stelzer, and Q.~Vanhellemont, ``Ocean color atmospheric correction methods in view of usability for different optical water types,'' \emph{Frontiers in Marine Science}, vol.~10, 2023. [Online]. Available: \url{https://www.frontiersin.org/articles/10.3389/fmars.2023.1129876}
\BIBentrySTDinterwordspacing

\bibitem{lee_2002}
Z.~Lee, K.~L. Carder, and R.~A. Arnone, ``Deriving inherent optical properties from water color: a multiband quasi-analytical algorithm for optically deep waters,'' \emph{Applied optics}, vol.~41, no.~27, pp. 5755--5772, 2002.

\bibitem{berthon_2004}
J.-F. Berthon and G.~Zibordi, ``Bio-optical relationships for the northern adriatic sea,'' \emph{International Journal of Remote Sensing}, vol.~25, no. 7-8, pp. 1527--1532, 2004.

\bibitem{Dicicco_2017}
\BIBentryALTinterwordspacing
A.~Di~Cicco, M.~Sammartino, S.~Marullo, and R.~Santoleri, ``Regional empirical algorithms for an improved identification of phytoplankton functional types and size classes in the mediterranean sea using satellite data,'' \emph{Frontiers in Marine Science}, vol.~4, 2017. [Online]. Available: \url{https://www.frontiersin.org/articles/10.3389/fmars.2017.00126}
\BIBentrySTDinterwordspacing

\bibitem{Volpe_2019}
\BIBentryALTinterwordspacing
G.~Volpe, S.~Colella, V.~E. Brando, V.~Forneris, F.~La~Padula, A.~Di~Cicco, M.~Sammartino, M.~Bracaglia, F.~Artuso, and R.~Santoleri, ``Mediterranean ocean colour level 3 operational multi-sensor processing,'' \emph{Ocean Science}, vol.~15, no.~1, pp. 127--146, 2019. [Online]. Available: \url{https://os.copernicus.org/articles/15/127/2019/}
\BIBentrySTDinterwordspacing

\bibitem{loisel2001seasonal}
H.~Loisel, E.~Bosc, D.~Stramski, K.~Oubelkheir, and P.-Y. Deschamps, ``Seasonal variability of the backscattering coefficient in the mediterranean sea based on satellite seawifs imagery,'' \emph{Geophysical Research Letters}, vol.~28, no.~22, pp. 4203--4206, 2001.

\bibitem{qiu2021relationships}
G.~Qiu, X.~Xing, E.~Boss, X.-H. Yan, R.~Ren, W.~Xiao, and H.~Wang, ``Relationships between optical backscattering, particulate organic carbon, and phytoplankton carbon in the oligotrophic south china sea basin,'' \emph{Optics Express}, vol.~29, no.~10, pp. 15\,159--15\,176, 2021.

\bibitem{morel1977analysis}
A.~Morel and L.~Prieur, ``Analysis of variations in ocean color 1,'' \emph{Limnology and oceanography}, vol.~22, no.~4, pp. 709--722, 1977.

\bibitem{neukermans2012situ}
G.~Neukermans, H.~Loisel, X.~M{\'e}riaux, R.~Astoreca, and D.~McKee, ``In situ variability of mass-specific beam attenuation and backscattering of marine particles with respect to particle size, density, and composition,'' \emph{Limnology and oceanography}, vol.~57, no.~1, pp. 124--144, 2012.

\bibitem{Boss2009IOPsSPM}
\BIBentryALTinterwordspacing
E.~Boss, L.~Taylor, S.~Gilbert, K.~Gundersen, N.~Hawley, C.~Janzen, T.~Johengen, H.~Purcell, C.~Robertson, D.~W.~H. Schar, G.~J. Smith, and M.~N. Tamburri, ``Comparison of inherent optical properties as a surrogate for particulate matter concentration in coastal waters,'' \emph{Limnology and Oceanography: Methods}, vol.~7, no.~11, pp. 803--810, 2009. [Online]. Available: \url{https://aslopubs.onlinelibrary.wiley.com/doi/abs/10.4319/lom.2009.7.803}
\BIBentrySTDinterwordspacing

\bibitem{datacmems}
``{E.U.} {C}opernicus {M}arine {S}ervice {I}nformation, https://doi.org/10.48670/moi-00299.''

\bibitem{Cmems}
\BIBentryALTinterwordspacing
``{C}opernicus {M}arine {S}ervice website.'' [Online]. Available: \url{https://data.marine.copernicus.eu/products}
\BIBentrySTDinterwordspacing

\bibitem{ioannidis2018intra}
E.~Ioannidis, C.~Lolis, C.~Papadimas, N.~Hatzianastassiou, and A.~Bartzokas, ``On the intra-annual variation of cloudiness over the mediterranean region,'' \emph{Atmospheric Research}, vol. 208, pp. 246--256, 2018.

\bibitem{candes_2010}
\BIBentryALTinterwordspacing
J.-F. Cai, E.~J. Cand\`{e}s, and Z.~Shen, ``A singular value thresholding algorithm for matrix completion,'' \emph{SIAM Journal on Optimization}, vol.~20, no.~4, pp. 1956--1982, 2010. [Online]. Available: \url{https://doi.org/10.1137/080738970}
\BIBentrySTDinterwordspacing

\bibitem{nguyen_2019}
L.~T. Nguyen, J.~Kim, and B.~Shim, ``Low-rank matrix completion: A contemporary survey,'' \emph{IEEE Access}, vol.~7, pp. 94\,215--94\,237, 2019.

\bibitem{OLIVER_2014}
\BIBentryALTinterwordspacing
M.~Oliver and R.~Webster, ``A tutorial guide to geostatistics: Computing and modelling variograms and kriging,'' \emph{CATENA}, vol. 113, pp. 56--69, 2014. [Online]. Available: \url{https://www.sciencedirect.com/science/article/pii/S0341816213002385}
\BIBentrySTDinterwordspacing

\bibitem{HOYER_2007}
\BIBentryALTinterwordspacing
J.~L. Høyer and J.~She, ``Optimal interpolation of sea surface temperature for the north sea and baltic sea,'' \emph{Journal of Marine Systems}, vol.~65, no.~1, pp. 176--189, 2007, marine Environmental Monitoring and Prediction. [Online]. Available: \url{https://www.sciencedirect.com/science/article/pii/S0924796306002910}
\BIBentrySTDinterwordspacing

\bibitem{Beckers_2006}
\BIBentryALTinterwordspacing
J.-M. Beckers, A.~Barth, and A.~Alvera-Azc{\'a}rate, ``Dineof reconstruction of clouded images including error maps -- application to the sea-surface temperature around corsican island,'' \emph{Ocean Science}, vol.~2, no.~2, pp. 183--199, 2006. [Online]. Available: \url{https://os.copernicus.org/articles/2/183/2006/}
\BIBentrySTDinterwordspacing

\bibitem{alvera_2009}
A.~Alvera-Azc{\'a}rate, A.~Barth, D.~Sirjacobs, and J.-M. Beckers, ``Enhancing temporal correlations in eof expansions for the reconstruction of missing data using dineof,'' \emph{Ocean Science}, vol.~5, no.~4, pp. 475--485, 2009.

\bibitem{Beauchamp2022}
M.~Beauchamp, Q.~Febvre, H.~Georgenthum, and R.~Fablet, ``4dvarnet-ssh: end-to-end learning of variational interpolation schemes for nadir and wide-swath satellite altimetry,'' \emph{Geoscientific Model Development Discussions}, pp. 1--37, 2022.

\bibitem{Barth2022}
A.~Barth, A.~Alvera-Azcárate, C.~Troupin, and J.-M. Beckers, ``Dincae 2.0: multivariate convolutional neural network with error estimates to reconstruct sea surface temperature satellite and altimetry observations,'' \emph{Geoscientific Model Development}, vol.~15, no.~5, pp. 2183--2196, 2022.

\bibitem{Wang2022}
Y.~Wang, X.~Zhou, Z.~Ao, K.~Xiao, C.~Yan, and Q.~Xin, ``Gap-filling and missing information recovery for time series of modis data using deep learning-based methods,'' \emph{Remote Sensing}, vol.~14, no.~19, 2022.

\bibitem{YUAN2020_DLRemoteSensing}
\BIBentryALTinterwordspacing
Q.~Yuan, H.~Shen, T.~Li, Z.~Li, S.~Li, Y.~Jiang, H.~Xu, W.~Tan, Q.~Yang, J.~Wang, J.~Gao, and L.~Zhang, ``Deep learning in environmental remote sensing: Achievements and challenges,'' \emph{Remote Sensing of Environment}, vol. 241, p. 111716, 2020. [Online]. Available: \url{https://www.sciencedirect.com/science/article/pii/S0034425720300857}
\BIBentrySTDinterwordspacing

\bibitem{Dorffer_IGARSS2024}
C.~Dorffer, T.~T.~N. Nguyen, R.~Fablet, and F.~Jourdin, ``Adaptive spatial and multi-variable generalization of 4dvarnet in ocean colour remote sensing,'' in \emph{IGARSS 2024 - 2024 IEEE International Geoscience and Remote Sensing Symposium}, 2024, pp. 7016--7019.

\bibitem{UNet2015}
O.~Ronneberger, P.~Fischer, and T.~Brox, ``U-net: Convolutional networks for biomedical image segmentation,'' \emph{International Conference on Medical image computing and computer-assisted intervention}, pp. 234--241, 2015.

\bibitem{Transformer2017}
A.~Vaswani, N.~Shazeer, N.~Parmar, J.~Uszkoreit, L.~Jones, A.~N. Gomez, L.~Kaiser, and I.~Polosukhin, ``Attention is all you need,'' in \emph{Advances in neural information processing systems}, vol.~30, 2017.

\bibitem{4DVarNetFablet2021Learning}
R.~Fablet, B.~Chapron, L.~Drumetz, E.~Memin, O.~Pannekoucke, and F.~Rousseau, ``Learning variational data assimilation models and solvers,'' \emph{Journal of Advances in Modeling Earth Systems}, vol.~13, no.~10, p. e2021MS002572, 2021.

\bibitem{cmems-mediterranean}
{CMEMS}, ``Ocean colour mediterranean biogeochemical layers, l3 product for the med. sea,'' \url{https://data.marine.copernicus.eu/product/OCEANCOLOUR_MED_BGC_L3_MY_009_143/description}, 2023.

\bibitem{Volpe_2018}
\BIBentryALTinterwordspacing
G.~Volpe, B.~B. Nardelli, S.~Colella, A.~Pisano, and R.~Santoleri, ``An operational interpolated ocean colour product in the mediterranean sea,'' \emph{New Frontiers in Operational Oceanography}, 2018. [Online]. Available: \url{https://api.semanticscholar.org/CorpusID:134745461}
\BIBentrySTDinterwordspacing

\bibitem{febvre2023training}
Q.~Febvre, J.~L. Sommer, C.~Ubelmann, and R.~Fablet, ``Training neural mapping schemes for satellite altimetry with simulation data,'' \emph{arXiv preprint arXiv:2309.14350}, 2023.

\bibitem{courtier1994strategy}
P.~Courtier, J.-N. Th{\'e}paut, and A.~Hollingsworth, ``A strategy for operational implementation of 4d-var, using an incremental approach,'' \emph{Quarterly Journal of the Royal Meteorological Society}, vol. 120, no. 519, pp. 1367--1387, 1994.

\bibitem{mikelsons2014destriping}
K.~Mikelsons, M.~Wang, L.~Jiang, and M.~Bouali, ``Destriping algorithm for improved satellite-derived ocean color product imagery,'' \emph{Optics express}, vol.~22, no.~23, pp. 28\,058--28\,070, 2014.

\bibitem{sauzede2016neuralBBP}
R.~Sauz{\`e}de, H.~Claustre, J.~Uitz, C.~Jamet, G.~Dall'Olmo, F.~d'Ortenzio, B.~Gentili, A.~Poteau, and C.~Schmechtig, ``A neural network-based method for merging ocean color and argo data to extend surface bio-optical properties to depth: Retrieval of the particulate backscattering coefficient,'' \emph{Journal of Geophysical Research: Oceans}, vol. 121, no.~4, pp. 2552--2571, 2016.

\bibitem{boss2008observations}
E.~Boss, D.~Swift, L.~Taylor, P.~Brickley, R.~Zaneveld, S.~Riser, M.~Perry, and P.~Strutton, ``Observations of pigment and particle distributions in the western north atlantic from an autonomous float and ocean color satellite,'' \emph{Limnology and Oceanography}, vol.~53, no. 5part2, pp. 2112--2122, 2008.

\end{thebibliography}
\vfill
\end{document}